\documentclass{emulateapj}
\usepackage[colorlinks,citecolor=blue]{hyperref}
\usepackage{amsmath}
\usepackage{graphicx}
\usepackage[caption=false]{subfig}

\shorttitle{Ion Aceleration by the Ion-cyclotron Instability}
\begin{document}
\title{Stochastic Ion Acceleration by the Ion-cyclotron Instability in a Growing Magnetic Field}  

\author{Francisco Ley\altaffilmark{1}, Mario Riquelme\altaffilmark{2}, Lorenzo Sironi\altaffilmark{3}, Daniel Verscharen\altaffilmark{4,5} \& Astor Sandoval\altaffilmark{6}}
\altaffiltext{1}{Department of Astronomy, University of Wisconsin-Madison; fley@astro.wisc.edu}
\altaffiltext{2}{Departamento de F\'isica, Facultad de Ciencias F\'isicas y Matem\'aticas, Universidad de Chile; mario.riquelme@dfi.uchile.cl}
\altaffiltext{3}{Department of Astronomy, Columbia University, New York, NY 10027 USA; lsironi@astro.columbia.edu}
\altaffiltext{4}{Mullard Space Science Laboratory, University College
London, Dorking, Surrey, UK;d.verscharen@ucl.ac.uk}
\altaffiltext{5}{Space Science Center, University of New Hampshire, Durham, NH 03824, USA}
\altaffiltext{6}{Instituto de Astrof\'isica, P. Universidad Cat\'olica de Chile; asandoval@astro.puc.cl}
 
\begin{abstract} 
\noindent  Using 1D and 2D particle-in-cell (PIC) simulations of a plasma with a growing magnetic field $\textbf{\textit{B}}$, we show that ions can be stochastically accelerated by the ion-cyclotron (IC) instability. As \textbf{\textit{B}} grows, an ion pressure anisotropy $p_{\perp,i} > p_{||,i}$ arises, due to the adiabatic invariance of the ion magnetic moment ($p_{||,i}$ and $p_{\perp,i}$ are the ion pressures parallel and perpendicular to \textbf{\textit{B}}). When initially $\beta_i = 0.5$ ($\beta_i \equiv 8\pi p_i/|\textbf{\textit{B}}|^2$, where $p_i$ is the ion isotropic pressure), the pressure anisotropy is limited mainly by inelastic pitch-angle scattering provided by the IC instability, which in turn produces a non-thermal tail in the ion energy spectrum. After \textbf{\textit{B}} is amplified by a factor $\sim 2.7$, this tail can be approximated as a power-law of index $\sim 3.4$ plus two non-thermal bumps, and accounts for $2-3\%$ of the ions and $\sim 18\%$ of their kinetic energy. On the contrary, when initially $\beta_i =2$, the ion scattering is dominated by the mirror instability and the acceleration is suppressed. This implies that efficient ion acceleration requires that initially $\beta_i \lesssim 1$. Although we focus on cases where $\textbf{\textit{B}}$ is amplified by plasma shear, we check that the acceleration occurs similarly if $\textbf{\textit{B}}$ grows due to plasma compression. Our results are valid in a sub-relativistic regime where the ion thermal energy is $\sim 10\%$ of the ion rest mass energy. This acceleration process can thus be relevant in the inner region of low-luminosity accretion flows around black holes.
\end{abstract} 


\section{Introduction}
\label{sec:intro}

\noindent Stochastic (or second-order Fermi) acceleration by plasma turbulence is considered a viable mechanism for producing non-thermal particles in several astrophysical environments. This process can in principle be driven by MHD plasma waves \citep[e.g.,][]{Chandran2003,ChoEtAl2006,LynnEtAl2014} and by kinetic plasma modes \citep[e.g.,][]{DermerEtAl1996,PetrosianEtAl2004}. In this work we use particle-in-cell (PIC) plasma simulations to show that ions can be stochastically accelerated by ion-cyclotron (IC) waves driven unstable in the presence of an ion pressure anisotropy with $p_{\perp,i} > p_{||,i}$ (where $p_{\perp,i}$ and $p_{||,i}$ are the ion pressures perpendicular and parallel to the local magnetic field \textbf{\textit{B}}, respectively).\newline

\noindent The condition $p_{\perp,i} \ne p_{||,i}$ is naturally expected in turbulent, weakly collisional plasmas. In these environments, Coulomb collisions are not able to break the adiabatic invariance of the magnetic moment $\mu_i$ of ions, which is defined as $\mu_i\equiv v_{\perp,i}^2/B$, where $v_{\perp,i}$ is the ion velocity perpendicular to $\textbf{\textit{B}}$ and $B=|\textbf{\textit{B}}|$. Thus, if $B$ grows (decreases), the conservation of $\mu_i$ will naturally produce a pressure anisotropy with $p_{\perp,i} > p_{||,i}$ ($p_{\perp,i} < p_{||,i}$). Examples of weakly collisional astrophysical plasmas where the condition $p_{\perp,i}\ne p_{\parallel,i}$ is possible are low-luminosity accretion disks around compact objects \citep[e.g.,][]{SharmaEtAl2006}, the intracluster medium \citep[ICM;][]{SchekochihinEtAl2005,Lyutikov2007}, and the heliosphere \citep[e.g.,][]{BaleEtAl2009,MarucaEtAl2011,VerscharenEtAl2019}.\newline

\noindent In these systems, the growth of ion pressure anisotropy is expected to be regulated by kinetic instabilities, which break the adiabatic invariance of $\mu_i$ via pitch-angle scattering of the ions. In the $p_{\perp,i}>p_{\parallel,i}$ regime, there are two relevant instabilities: the mirror and the ion-cyclotron (IC) instabilities. The mirror instability consists of non-propagating, compressional modes, with their dominant modes having wave vectors $\textbf{\textit{k}}$ oblique to the direction of $\textbf{\textit{B}}$ \citep{Hasegawa1969, SouthwoodEtAl1993}. The IC instability, on the other hand, consists of propagating electromagnetic modes, with their dominant waves having $\textbf{\textit{k}} \parallel \textbf{\textit{B}}$ \citep{AndersonEtAl1991, Gary1992}. Whether the ion pitch-angle scattering is dominated by the IC or mirror instability essentially depends on how fast the instabilities grow for a given plasma regime. In this work we show that, in a regime dominated by the IC instability, significant non-thermal ion acceleration can occur due to scattering by the IC waves.\newline

\noindent Our study will use particle-in-cell (PIC) plasma simulations to study a homogeneous plasma in which ion and electron pressure anisotropies are self-consistently produced by the continuous growth of a background magnetic field $B$. In our simulations, the magnetic field will grow on time scales significantly longer than the initial, exponential growth regime of the mirror and IC instabilities. This will allow us to capture the long-term, saturated state of the instabilities, which should be the dominant regime in astrophysical systems where $B$ experiences significant amplifications. In most of our simulations, the magnetic field will be amplified through imposing a slow shear motion in the plasma, which will increase $B$ due to magnetic flux conservation. However, we will also use simulations of compressing plasmas to show that our main results are fairly independent of the specific mechanism that drives the growth of $B$. In this study we focus on conditions applicable to the inner regions of low-luminosity accretion flows around black holes. This will be done by assuming in all of our runs a hot plasma with initially equal ion and electron temperatures ($T_i=T_e$) and with $k_BT_i/m_ic^2=0.05$ ($k_B$ is the Boltzmann constant, $m_i$ is the mass of the ions, and $c$ is the speed of light).  \newline

\noindent Our paper is organized as follows. In \S \ref{sec:numsetup} we present our simulation method and setup. In \S \ref{sec:ionaccel} we use 2D simulations to show that ions can be accelerated by the IC instability under the condition that initially $\beta_i \lesssim 1$ ($\beta_i \equiv 8\pi p_i/|\textbf{\textit{B}}|^2$, where $p_i$ is the ion isotropic pressure). In \S \ref{sec:1D} we use 1D simulations to clarify the role of the IC and mirror modes in the acceleration, as well as to show that our results are independent of the rate at which $B$ is amplified and of the numerical ion to electron mass ratio, $m_i/m_e$. In \S \ref{sec:nature} we describe the acceleration in further details, connecting the growth of IC modes of different wavenumber $k$ with the acceleration of ions of different energy. In \S \ref{sec:comp} we show that our results are fairly independent on whether $B$ is amplified via plasma shear or compression. In \S \ref{sec:conclu} we summarize our results and present our conclusions. Additionally, in Appendix \S \ref{1druns} we provide details on the implementation of our 1D simulations, and in Appendix \S \ref{linear} we use linear theory to analyze the applicability of our simulation results to realistic astrophysical environments.   

\section{Simulation Setup}
 \label{sec:numsetup}
 \begin{figure}[t!]  
\centering 
\vspace*{-0.1cm}
\includegraphics[width=8cm]{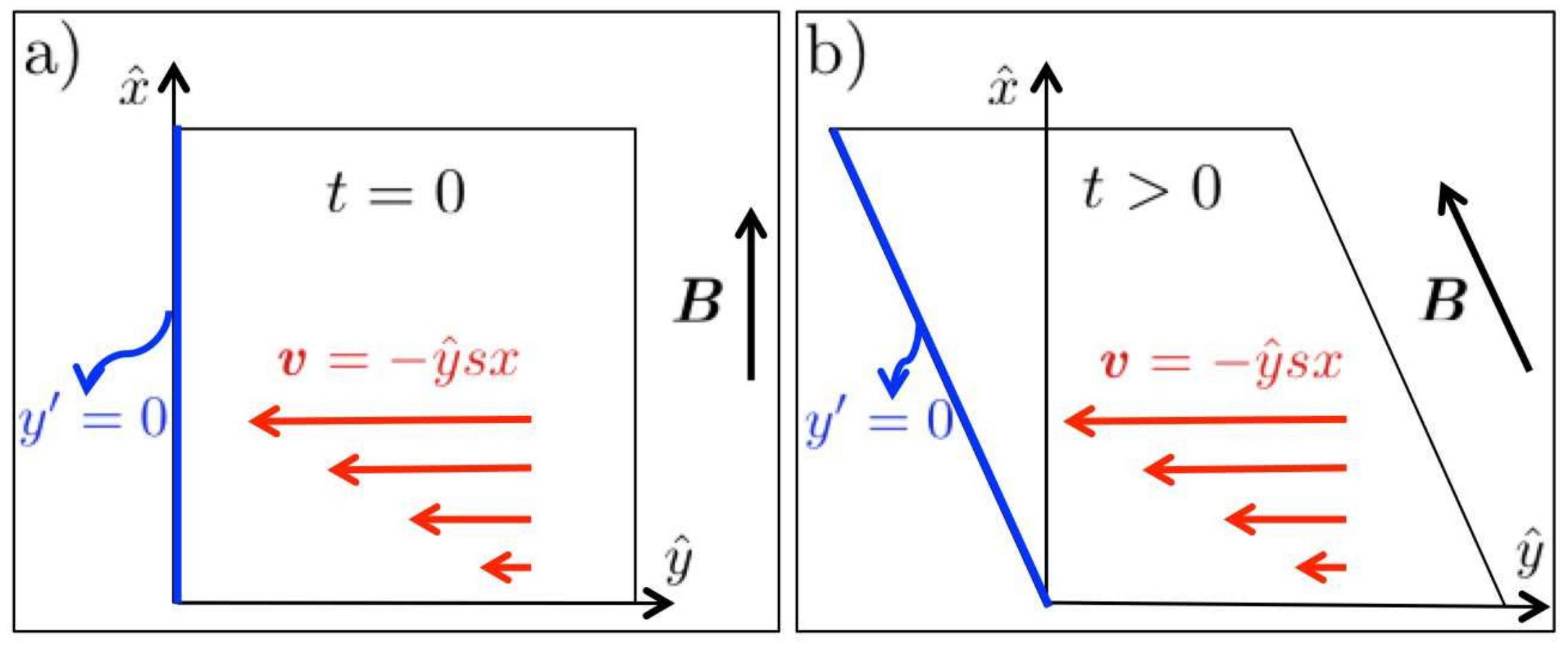}
\caption{\small Panels $a$ and $b$ show a sketch of the simulation domain in our 2D shearing simulations at $t=0$ and $t> 0$, respectively. The 2D domain follows the shearing flow of the plasma (red arrows), acquiring a parallelogram shape. Magnetic flux conservation changes the magnitude and orientation of the background magnetic field $\textbf{\textit{B}}$, which is always parallel to the non-horizontal sides of the parallelogram. The blue lines show the domain of our 1D runs.} 
\label{fig:shearscheme} 
\end{figure} 
\begin{deluxetable}{llllll} \tablecaption{Parameters of the simulations} \tablehead{ \colhead{Runs}&\colhead{$\omega_{c,i}^{\textrm{init}}/s$}& $m_i/m_e$ &\colhead{$\beta_i^{\textrm{\textrm{init}}}$}&\colhead{N$_{\textrm{ppc}}$}&\colhead{L/R$_{L,i}^{\textrm{init}}$[D]} \\
&(or $\omega_{c,i}^{\textrm{init}}/q$)&&&&} \startdata
  S2m2b0.5 & 800 & 2 & 0.5 &160 & 60[2]\\
  S2m2b2 & 800 & 2 & 2 & 160& 60[2] \\
  S2m10b0.5 & 800 & 10 & 0.5 &160 & 60[2]\\
  S2m10b2 & 800 & 10 & 2 & 160& 60[2] \\
  S1m2b0.5 & 800 & 2 & 0.5 & 640& 150[1] \\
  S1m2b2 & 800 & 2 & 2 & 640& 150[1] \\
  S1m8b0.5 & 800 & 8 & 0.5 & 640& 150[1] \\
  S1m32b0.5 & 800 & 32 & 0.5 & 640& 150[1] \\
  S1m128b0.5 & 800 & 128 & 0.5 & 1280& 150[1]\\
  S1m8b0.5b & 400 & 8 & 0.5 & 640& 150[1] \\
  S1m8b0.5c & 1600 & 8 & 0.5 & 640& 150[1] \\
  S1m8b0.5d & 3200 & 8 & 0.5 & 3600& 150[1] \\
  C1m8b0.5a & 1600 & 8 & 0.5  & 3600 & 190[1] \\
  C1m8b0.5b & 3200 & 8 & 0.5  & 3600 & 190[1] \\
  C1m16b0.5a & 1600 & 16 & 0.5  & 3600 & 190[1] \\
  C1m16b0.5b & 3200 & 16 & 0.5  & 3600 & 190[1]
\enddata \tablecomments{Simulation parameters: the initial ion cyclotron frequency $\omega_{c,i}^{\textrm{init}}$ (in units of $s$ in the shearing runs and of $q$ in the compressing runs), $m_i/m_e$, $\beta_{i}^{\textrm{init}}$, the number of particles per cell N$_{\textrm{ppc}}$ (considering ions and electrons), and the initial box size in units of the initial ion Larmor radius $L/R_{L,i}^{\textrm{\textrm{init}}}$ (with the number of dimensions D in squared parenthesis). In 2D, $L$ corresponds to the height and width of the box. In all runs initially $k_BT_i/m_ic^2=0.05$, $T_e=T_i$, the electron skin depth $c/\omega_{p,e}/\Delta_x=15$ (where $\Delta_x$ is the grid point separation), the speed of light $c=0.225 \Delta_x/\Delta_t$ (shearing runs) and 0.15$\Delta_x/\Delta_t$ (compressing runs), where $\Delta_t$ is the simulation time step.} \label{table} \end{deluxetable}
\noindent We use the PIC code TRISTAN-MP \citep{Buneman93, Spitkovsky05} to simulate both a shearing and a compressing plasma made of ions and electrons. In the shearing case, the plasma is initially in presence of a homogeneous initial magnetic field that points along the $x$ axis, $\textbf{\textit{B}}=B_0 \hat{x}$. This field is amplified by imposing a shear plasma velocity $\textbf{\textit{v}} = -sx\hat{y}$ (represented by red arrows in Fig. \ref{fig:shearscheme}$a$), where $x$ is the distance along $\hat{x}$ and $s$ is the shear rate. This way the background magnetic field $\textbf{\textit{B}}$ in the simulation permanently increases and changes direction due to magnetic flux conservation, with its $y$-component evolving as $dB_y/dt = -sB_0$, while $d B_x/dt = dB_z/dt =0$ (Fig. \ref{fig:shearscheme}$b$ shows how $\textbf{\textit{B}}$ changes orientation for $t>0$). Due to $\mu_j$ conservation, this magnetic growth drives $p_{\perp,j} > p_{||,j}$ during the whole simulation, allowing the triggering of kinetic instabilities that limit the pressure anisotropies.\newline 

\noindent Our 2D shearing runs use initially square simulation domains (as the one depicted in Fig. \ref{fig:shearscheme}$a$) that follow the mean shear motion of the plasma. Therefore, the 2D domain acquires a parallelogram shape for $t > 0$ (as shown in Fig. \ref{fig:shearscheme}$b$). The positions of the plasma particles are therefore given in terms of the so called `shearing coordinates', which are described both in Appendix \ref{1druns} of this paper as well as in the Appendix of \cite{RiquelmeEtAl2012}. In our 1D shearing runs, on the other hand, the simulation domain corresponds to the blue narrow stripe shown in Figs. \ref{fig:shearscheme}$a$ and \ref{fig:shearscheme}$b$, which also moves with the shearing flow. Since the symmetry axis of this 1D domain is permanently parallel to $\textbf{\textit{B}}$, our 1D approach allows to capture waves that propagate parallel to $\textbf{\textit{B}}$. The self-consistent implementation of the 1D runs requires a small change in the definition of our shear coordinates, which is explained in detail in Appendix \ref{1druns}. In \S \ref{1d2dcompa} we show that our 1D runs give essentially the same results as our 2D runs as long as the dominant instabilities produce modes parallel to $\textbf{\textit{B}}$, such as the IC instability.\newline

\noindent In our compressing plasma runs, on the other hand, the simulation box is compressed along the two directions perpendicular to the background field $\textbf{\textit{B}}$, producing both the permanent growth of $\textbf{\textit{B}}$ and of $p_{\perp,j}/p_{\parallel,j}$. For this we use the same setup as in \cite{SironiEtAl2015}. In this case, $\textbf{\textit{B}}$ evolves as $\textbf{\textit{B}}=\hat{x} B_0/(1+qt)^2$, where the constant $q$ provides the time scale for the plasma compression.\newline       

\noindent Our plasma parameters are the initial temperature of ions and electrons ($T_i$ and $T_e$), the initial ratio between ion pressure and magnetic pressure ($\beta_i^{\textrm{init}}$), the ion to electron mass ratio $m_i/m_e$, and the ion ``magnetization", which is defined as the ratio between the initial ion cyclotron frequency ($\omega_{c,i}^{\textrm{init}}$) and $s$ (for shearing plasma runs) or $q$ (for compressing plasma runs). The initial ion cyclotron frequency is defined as $\omega_{c,i}^{\textrm{init}}=eB_0/m_ic$, with $e$ and $B_0$ being the magnitude of the electron and ion electric charges and the initial magnetic field. \newline

\noindent As mentioned above, all of our shearing and compressing simulations have initially $T_i=T_e$ and $k_BT_i/m_ic^2 = 0.05$. Also, these runs use $m_i/m_e$ and ion magnetizations much smaller than expected in real astrophysical settings.\footnote{ For instance, at $\sim 10$ Schwarzschild radii from the super-massive black hole Sgr A*, one expects $\omega_{c,i}/s \sim 10^{8}$ \citep[e. g.,][]{PontiEtAl2017}, where we have approximated $s$ as the Keplerian angular frequency at that radius.} Because of this, the dependence of the ion acceleration on these parameters will be carefully assessed. (With our main conclusion being that neither $m_i/m_e$ or $\omega_{c,i}^{\textrm{init}}/s$ (or $\omega_{c,i}^{\textrm{init}}/q$) play a significant role.)\newline

\noindent The numerical parameters in our runs are: the number of macro-particles per cell (N$_{\textrm{ppc}}$), the electron skin depth in terms of grid point spacing ($c/\omega_{p,e}/\Delta_x$, where $\omega_{p,e}^2=4\pi n_e e^2/m_e$ is the electron plasma frequency and $n_e$ is the electron number density), and the box size in terms of the initial ion Larmor radius ($L/R_{L,i}^{\textrm{\textrm{init}}}$; $R_{L,i}^{\textrm{\textrm{init}}} = v_{th,i}/\omega_{c,i}^{\textrm{init}}$, where $v_{th,i}^2=k_BT_i/m_i$). Table \ref{table} shows a summary of our key simulations. We ran a series of simulations ensuring that the numerical parameters do not affect our results. The runs used just for numerical convergence are not in Table \ref{table}.\newline 
   
\begin{figure}
\hspace*{-0.7cm}
\includegraphics[width=.55\textwidth]{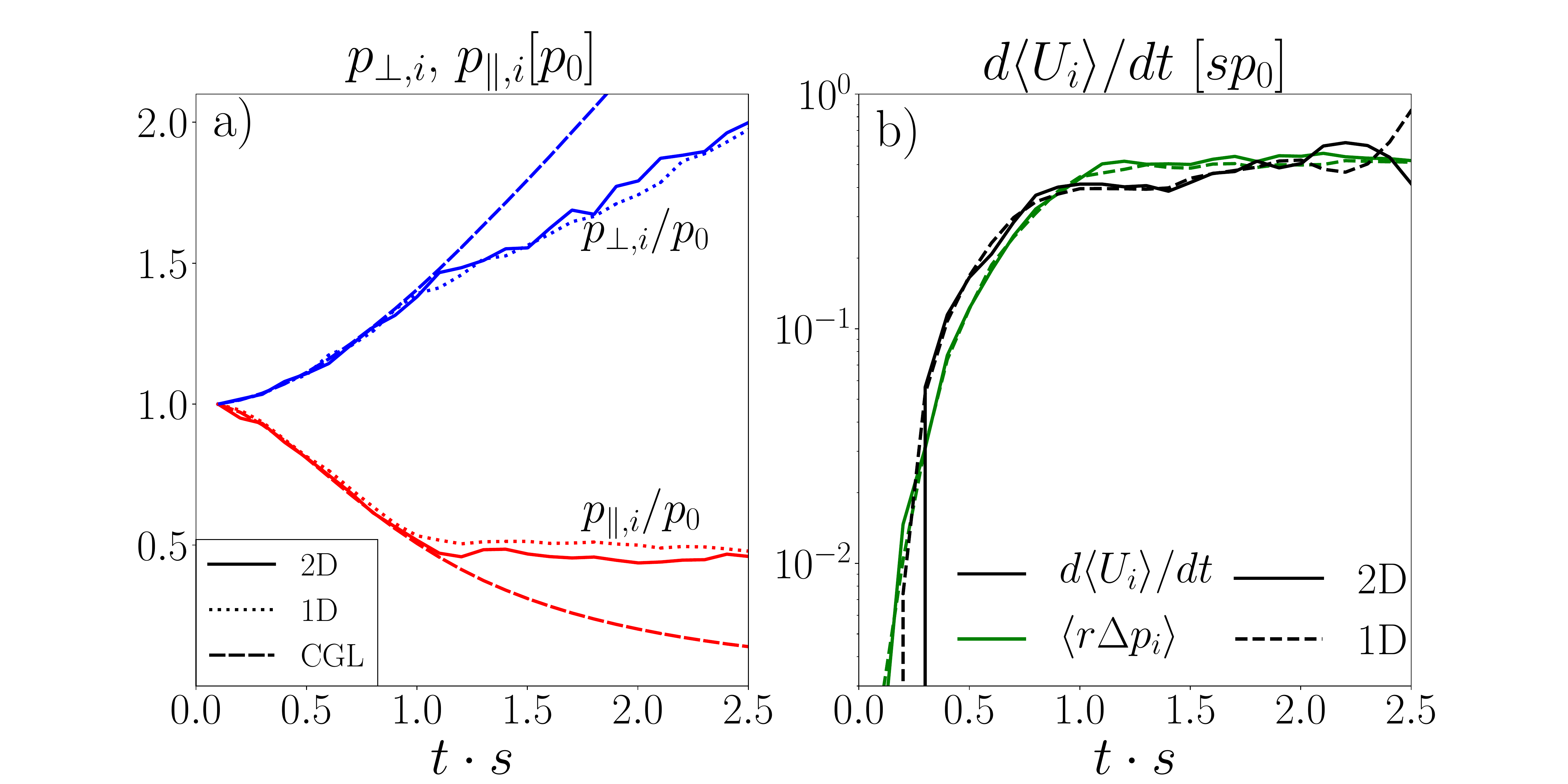}
\vspace*{-0.4cm}
\caption{\small Panel $a$: $p_{i,\parallel}$ (red) and $p_{i,\perp}$ (blue) for 2D run S2m2b0.5 (solid) and 1D run S1m2b0.5 (dotted), both with $m_i/m_e=2$ and $\beta_i^{\textrm{init}}=0.5$. The dashed lines show the corresponding double-adiabatic behavior \citep{ChewEtAl1956}. Panel $b$: the volume-averaged derivative of the ion internal energy density $dU_i/dt$ (black) and the ``anisotropic viscosity" prediction $r\Delta p_i$ (green; see Equation \ref{anvis}) for the same 2D and 1D runs S2m2b0.5 (solid) and S1m2b0.5 (dashed). Both $dU_i/dt$ and $r\Delta p_i$ evolve very similarly in 1D and 2D, and in both cases the ion energy gain reproduces fairly well the anisotropic viscosity prediction.}
\label{fig:cgl}
\end{figure}
\begin{figure}
 \vspace*{-0.4cm}
\subfloat{
  \centering
\hspace*{-0.3cm}
  \includegraphics[width=0.5\textwidth]{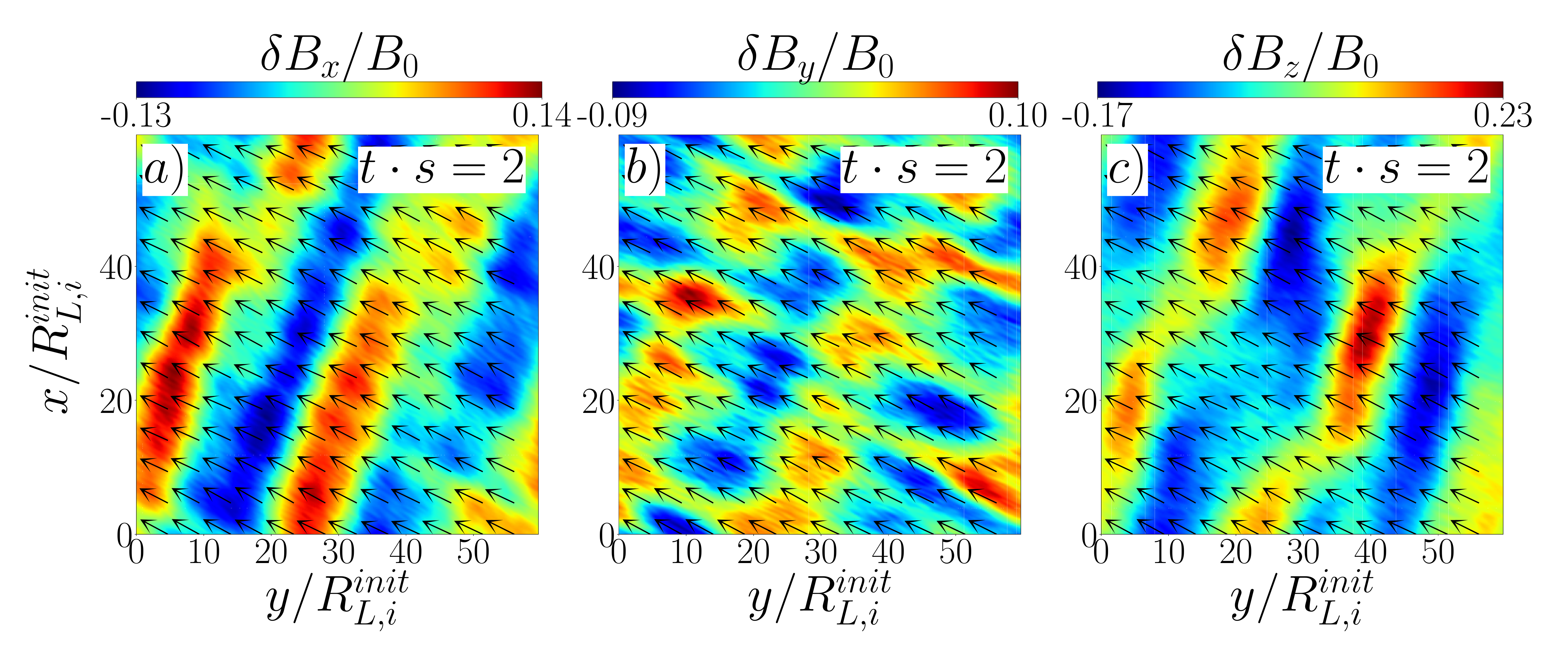}
}
\vspace*{-0.6cm}
\\
\subfloat{
  \centering
 \hspace*{-0.3cm}
  \includegraphics[width=0.5\textwidth]{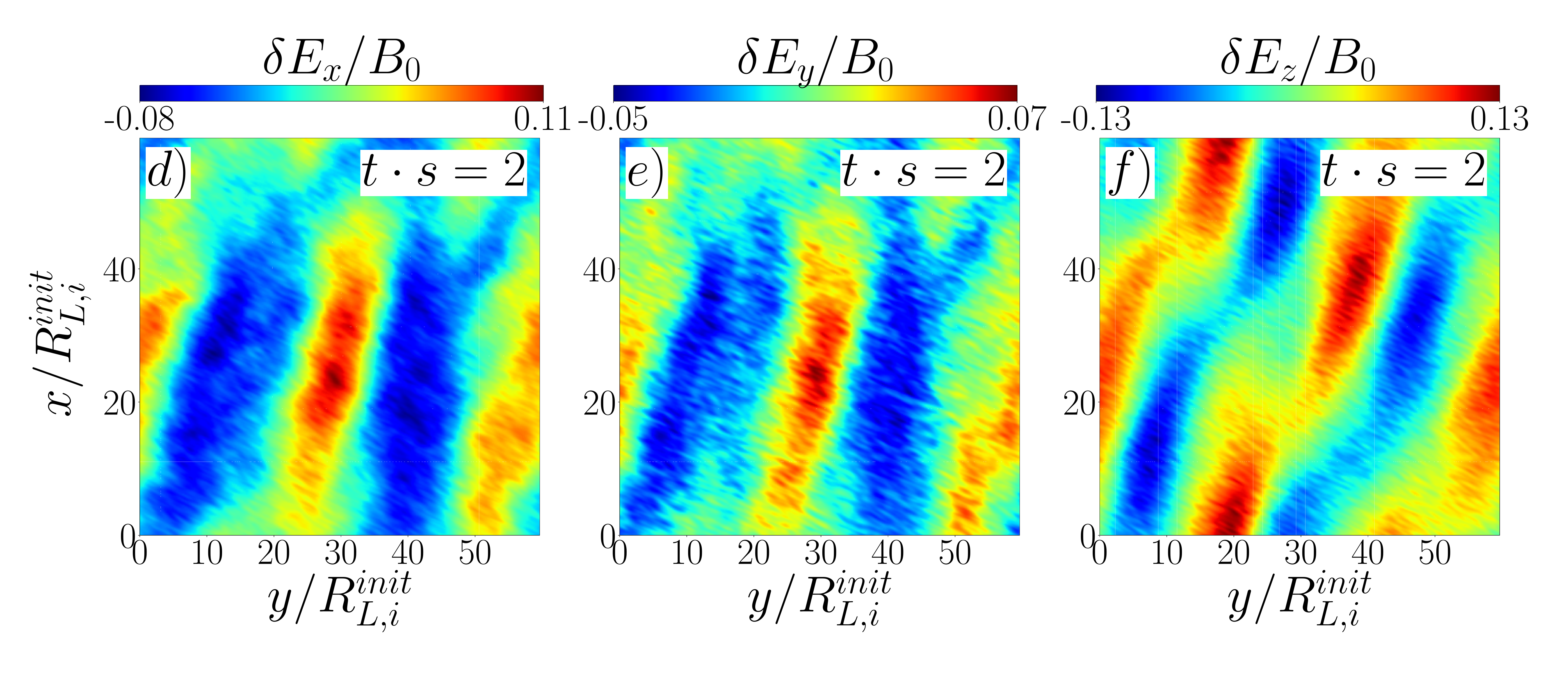}
}
\vspace*{-0.3cm}
\caption{\small Panels $a$, $b$ and $c$: the $x$, $y$, and $z$ components of the magnetic fluctuation $\delta \textbf{\textit{B}}$ for run S2m2b0.5 at $t\cdot s=2$. The black arrows show the direction of the total magnetic field $\textbf{\textit{B}}$. Panels $d$, $e$ and $f$: the same components but for the electric field fluctuation $\delta \textbf{\textit{E}}$.}
\label{fig:fields}
\end{figure}

\section{Ion Acceleration by the IC Instability}
\label{sec:ionaccel}
\noindent We use 2D, shearing plasma simulations to show that ions can be stochastically accelerated by the IC modes. First, we show the example of runs dominated by the IC instability (with $\beta_{i}^{\textrm{init}}=0.5$), demonstrating that in this case a prominent non-thermal tail appears. Then, using simulations with $\beta_{i}^{\textrm{init}}=2$, we show that for $\beta_{i}^{\textrm{init}}\gtrsim 1$ the mirror instability dominates, with a corresponding suppression of the accelerating effect of the IC modes. We use 2D simulations with $m_i/m_e=2$ and 10 to show that, as long as the ion physics is concerned, our results are fairly independent of the value of $m_i/m_e$. 
\subsection{IC vs. mirror dominated regimes} 
\label{sec:icregime}
\noindent Figure \ref{fig:cgl}$a$ shows in blue-solid and red-solid lines the respective evolutions of $p_{\perp,i}$ and $p_{\parallel,i}$ for run S2m2b0.5, which uses $\beta_i^{\textrm{init}}=0.5$, $m_i/m_e=2$ and $\omega_{c,i}^{\textrm{init}}/s=800$. We see that until $t\cdot s \approx 1$ the evolutions of both $p_{\perp,i}$ and $p_{\parallel,i}$ are in agreement with the ``double adiabatic" prediction \citep[dashed lines;][]{ChewEtAl1956}, which is due to the conservation of $\mu_i$ and of the second adiabatic invariant. At $t\cdot s \gtrsim 1$, the adiabatic evolution of $p_{\perp,i}$ and $p_{\parallel,i}$ is broken by the appearance of ion pressure anisotropy instabilities, which produce rapid pitch-angle scattering of the ions. \newline    

\noindent We see from Figure \ref{fig:fields} that these instabilities are dominated by IC modes. Indeed, Figs. \ref{fig:fields}$a$, \ref{fig:fields}$b$ and \ref{fig:fields}$c$ show a snapshot at $t\cdot s=2$ of the three components of $\delta \textbf{\textit{B}}$ (where $\delta \textbf{\textit{B}} \equiv \textbf{\textit{B}} - \langle \textbf{\textit{B}} \rangle$ and $\langle \rangle$ denotes an average over the entire box volume). Considering that the black arrows represent the direction of $\langle \textbf{\textit{B}} \rangle$, we see that $\delta \textbf{\textit{B}}$ is dominated by nearly parallel modes, which mainly appear in $\delta B_x$ and $\delta B_z$. This is indeed consistent with the presence of transverse, circularly polarized IC modes. $\delta B_y$ shows a mixture of the nearly parallel modes, plus subdominant oblique modes, which are consistent with the presence of mirror modes. Indeed, these modes mainly contribute to the $\delta \textbf{\textit{B}}$ components parallel to the plane of the simulation (they show no $\delta B_z$ component), which is in line with the expectation that $\delta \textbf{\textit{B}}$ of the mirror modes is nearly perpendicular to $\textbf{\textit{k}} \times \textbf{\textit{B}}$ \citep{PokhotelovEtAl2004}.\newline

\begin{figure}[t!]
\hspace*{-0.4cm}
\vspace*{-0.7cm}
\includegraphics[width=.52\textwidth]{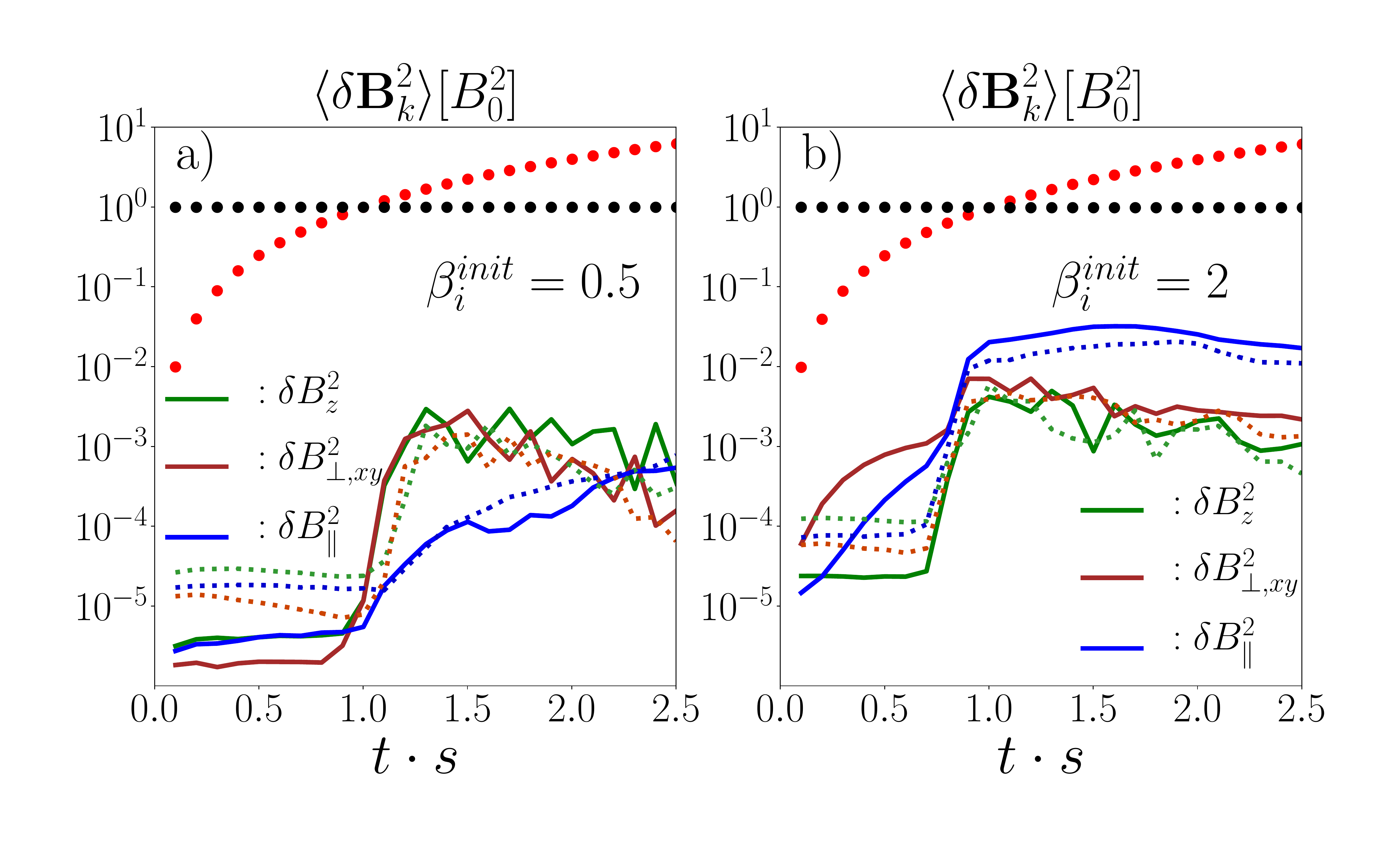}
\caption{\small Panels $a$ and $b$: in solid lines the volume-averaged magnetic energy of $\delta \textbf{\textit{B}}$ along different axes and as a function of time for runs S2m2b0.5 ($m_i/m_e=2$ and $\beta_i^{\textrm{init}}=0.5$; panel $a$) and S2m2b2 ($m_i/m_e=2$ and $\beta_i^{\textrm{init}}=2$; panel $b$). $\delta B_{\parallel}$ (blue) is the component parallel to $\langle \textbf{\textit{B}} \rangle$. $\delta B_{xy,\perp}$ (red) and $\delta B_{z}$ (green) are the components perpendicular to $\langle \textbf{\textit{B}} \rangle$ but, respectively, parallel and perpendicular to the simulation plane. The thick dotted-black and dotted-red lines represent the volume-averaged values of $\langle B_x \rangle^2$ and $\langle B_y\rangle ^2$, respectively, which show how $\langle \textbf{\textit{B}} \rangle$ grows during the simulation. The thin dotted lines show the same quantities but for runs S2m10b0.5 ($m_i/m_e=10$ and $\beta_i^{\textrm{init}}=0.5$; panel $a$) and S2m10b2 ($m_i/m_e=10$ and $\beta_i^{\textrm{init}}=2$; panel $b$).}
\label{fig:evolb}
\end{figure}

\noindent The nearly parallel IC modes can also be seen from Figs. \ref{fig:fields}$d$, \ref{fig:fields}$e$ and \ref{fig:fields}$f$, which show the fluctuations in the electric field, $\delta \textbf{\textit{E}}$.\footnote{Since our simulations are performed in the ``shearing coordinate" frame \citep{RiquelmeEtAl2012}, there is not an electric field associated to the large-scale shearing motion of the plasma. Therefore, $\delta \textbf{\textit{E}}$ corresponds to the entire electric field present in the simulation.} This electric field is expected since the IC modes have finite phase velocities, $v_{\phi}$, which is related to $\delta \textbf{\textit{E}}$ by $|\delta \textbf{\textit{E}}| = |\delta \textbf{\textit{B}}|v_{\phi}/c$ (this is a consequence of Faraday's law applied to the transverse IC modes). The mirror modes, on the other hand, are `purely growing' \citep[see, e.g.,][]{SouthwoodEtAl1993}, which means that their phase velocity vanishes. This implies that, as we see in Fig. \ref{fig:fields}, no electric field associated to the subdominant mirror modes should be present.\newline

\noindent The dominance of the IC modes can also be seen from Figure \ref{fig:evolb}$a$, which shows in solid lines the magnetic energy of $\delta \textbf{\textit{B}}$ along different axes as a function of time for run S2m2b0.5. This energy is expressed in terms of the $\delta \textbf{\textit{B}}$ components parallel to $\langle \textbf{\textit{B}} \rangle$ ($\delta B_{\parallel}$; solid-blue), perpendicular to $\langle \textbf{\textit{B}} \rangle$ but parallel to the plane of the simulation ($\delta B_{xy,\perp}$; solid-red), and perpendicular to both $\langle \textbf{\textit{B}} \rangle$ and the plane of the simulation ($\delta B_{z}$; solid-green). During most of the simulation, the energy of the magnetic fluctuations is indeed contained mainly in $\delta B_{xy,\perp}$ and $\delta B_{z}$, implying that the IC modes have the largest amplitude during most of the simulation time. By the end of the run ($t\cdot s=2.5$), however, $\delta B_{\parallel}^2$ becomes comparable to $\delta B_{xy,\perp}^2$ and $\delta B_{z}^2$, implying that in the long term the mirror fluctuations can still reach amplitudes comparable to the IC modes. We have thus decided to concentrate on the regime where the IC instability clearly dominates the pitch-angle scattering of the ions by running the simulations until $t\cdot s=2.5$ (thus with a maximum $B$ amplification factor of $\sim 2.7$).\newline

\noindent  Figure \ref{fig:evolb}$b$, on the other hand, shows the evolution of the same magnetic energy components for the 2D run S2m2b2 ($m_i/m_e=2$ and $\beta_i^{\textrm{init}}=2$). In this run, the ions are under the same conditions as in run S2m2b2, but with a smaller initial background magnetic field so that $\beta_i^{\textrm{init}}=2$. We see that in this case $\delta B_{z}$ and $\delta B_{xy,\perp}$ are subdominant in the saturated stage of the instabilities, and the energy in the magnetic fluctuations is dominated by $\delta B_{\parallel}^2$. This result indicates that the oblique mirror modes are more prominent than the IC modes in this case, with the transition from IC-dominated to mirror-dominated regimes happening at $\beta_i^{\textrm{init}}\sim 1$. \newline

\noindent In order to explore the sensitivity of this transition to $m_i/m_e$, in Figs. \ref{fig:evolb}$a$ and \ref{fig:evolb}$b$ we overplot $\delta B_{\parallel}^2$, $\delta B_{xy,\perp}$, and $\delta B_{z}$ for simulations S2m10b0.5 and S2m10b2, which have the same ion conditions as in runs S2m2b0.5 and S2m2b2 (i.e., the same values of $\beta_i^{\textrm{init}}$, $k_BT_i/m_ic^2$ and $\omega_{c,i}^{\textrm{init}}/s$), but with $m_i/m_e=10$ instead of $m_i/m_e=2$. We see that for the two $\beta_i^{\textrm{init}}$ the evolutions of $\delta B_{xy,\perp}$, $\delta B_z$, and $\delta B_{\parallel}$ are fairly independent of $m_i/m_e$. Thus the mass ratio does not appear to affect significantly the dominance of the IC instability for $\beta_i^{\textrm{init}} \lesssim 1$. \newline

\noindent Additionally, in Appendix \S \ref{linear} we use linear theory calculations to show that the condition $\beta_i^{\textrm{init}} \lesssim 1$ for the dominance of the IC instability should continue to hold even in realistic astrophysical plasma conditions, with $m_i/m_e=1836$, $\omega_{c,i}^{\textrm{init}}/s \gg 800$ and $T_e=T_i$. In the next section we show that this IC dominance also results in a significant non-thermal ion acceleration, which is strongly suppressed when the mirror modes dominate.
 
\subsection{Ion heating and acceleration}
\label{sec:ionspectic}

\noindent It is well known that in a collisionless, shearing plasma the particles are heated by the so called ``anisotropic viscosity". Indeed, for a homogeneous plasma subject to shear, the internal energy density for species $j$, $U_j$, evolves as \citep{Kulsrud1983, SnyderEtAl1997}
\begin{equation}
\frac{dU_j}{dt} = r\Delta p_j,
\label{anvis}
\end{equation}
where $r$ is the growth rate of the magnetic field ($r \equiv (dB/dt)/B$). Equation \ref{anvis} is fairly well reproduced in our simulations, as can be seen from Figure \ref{fig:cgl}$b$, which shows the volume-averaged heating rate of ions $dU_i/dt$ (solid-black) and $r\Delta p_i$ (solid-green) for run S2m2b0.5.\newline 
\begin{figure}[t!]
 \hspace*{-0.3cm} 
 \vspace*{-0.2cm}
\includegraphics[width=0.52\textwidth]{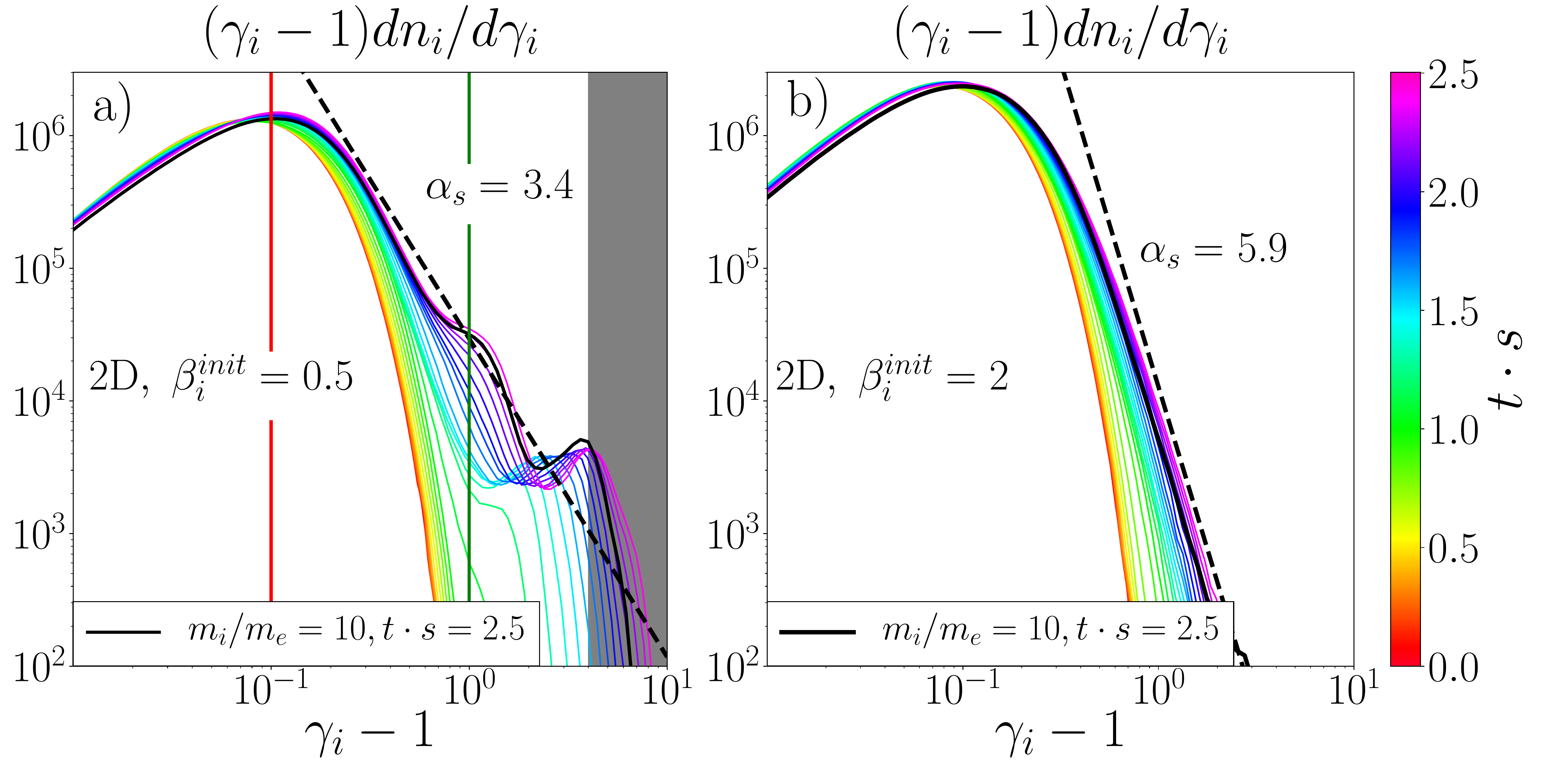}
\caption{\small Panels $a$ and $b$: evolutions of the ion spectra for runs S2m2b0.5 ($m_i/m_e=2$, $\beta_{i}^{\textrm{init}}=0.5$) and S2m2b2 ($m_i/m_e=2$, $\beta_{i}^{\textrm{init}}=2$), which correspond to regimes dominated by the IC and mirror instabilities, respectively. The color bar indicates the time for each spectrum. The overplotted black lines correspond to the final spectra of runs S2m10b0.5 ($m_i/m_e=10$, $\beta_{i}^{\textrm{init}}=0.5$; panel $a$) and S2m10b2 ($m_i/m_e=10$, $\beta_{i}^{\textrm{init}}=2$; panel $b$), which, apart from using different mass ratios, assume the same plasma conditions as runs S2m2b0.5 and S2m2b2, respectively.}
\label{fig:spectra}
\end{figure}

\noindent Since $dU_i/dt$ is dominated by the ion pressure anisotropy, the ion heating is ultimately regulated by the pitch-angle scattering provided by either the IC or mirror instabilities. In this section we show that when this scattering is provided mainly by the IC modes, it can also give rise to significant stochastic ion acceleration.\newline

\noindent Figure \ref{fig:spectra}$a$ shows the evolution of the ion spectrum for 
run S2m2b0.5 ($\beta_{i}^{\textrm{init}}=0.5$, $m_i/m_e=2$), with the color bar indicating the time for each spectrum. This simulation shows the rapid growth of a non-thermal tail that starts once the IC instability grows and saturates ($t\cdot s\sim 1$, as seen from Figure \ref{fig:evolb}$a$). By $t\cdot s=2.5$ the tail can be approximated by a power law $dn_i/d\gamma_i \propto (\gamma_i-1)^{-\alpha_s}$ with spectral index $\alpha_s \approx 3.4$ plus two bumps ($\gamma_i$ is the ion Lorentz factor). The non-thermal tail at $t\cdot s=2.5$ reaches Lorentz factors $\gamma_i \sim 10$, and contains $\sim 2-3\%$ of the ions and $\sim 18\%$ of their energy.\footnote{After fitting the low energy part of $dn_i/d\gamma_i$ to a thermal Maxwell-Boltzmann distribution, we define the non-thermal tail through the condition that $dn_i/d\gamma_i$ is at least a factor $2$ larger than the expectation for the thermal distribution.} The solid-black line in Fig. \ref{fig:spectra}$a$ represents the final ($t\cdot s=2.5$) spectra for the analogous run S2m10b0.5 (with $m_i/m_e=10$ instead of 2). The small difference between the $m_i/m_e=2$ and 10 cases shows that, as long as the ion parameters $k_BT_i/m_ic^2$, $\beta_{i}^{\textrm{init}}$, and $\omega_{c,i}^{\textrm{init}}/s$ are the same, the ion to electron mass ratio does not play a significant role in determining the ion acceleration efficiency. The independence of the acceleration mechanism on $m_i/m_e$ as well as on $\omega_{c,i}^{\textrm{init}}/s$ will be further tested using 1D simulations in \S \ref{sec:1D}.\newline
\begin{figure}[t!] 
\hspace*{-0.3cm}
\vspace*{-0.2cm}
\includegraphics[width=0.52\textwidth]{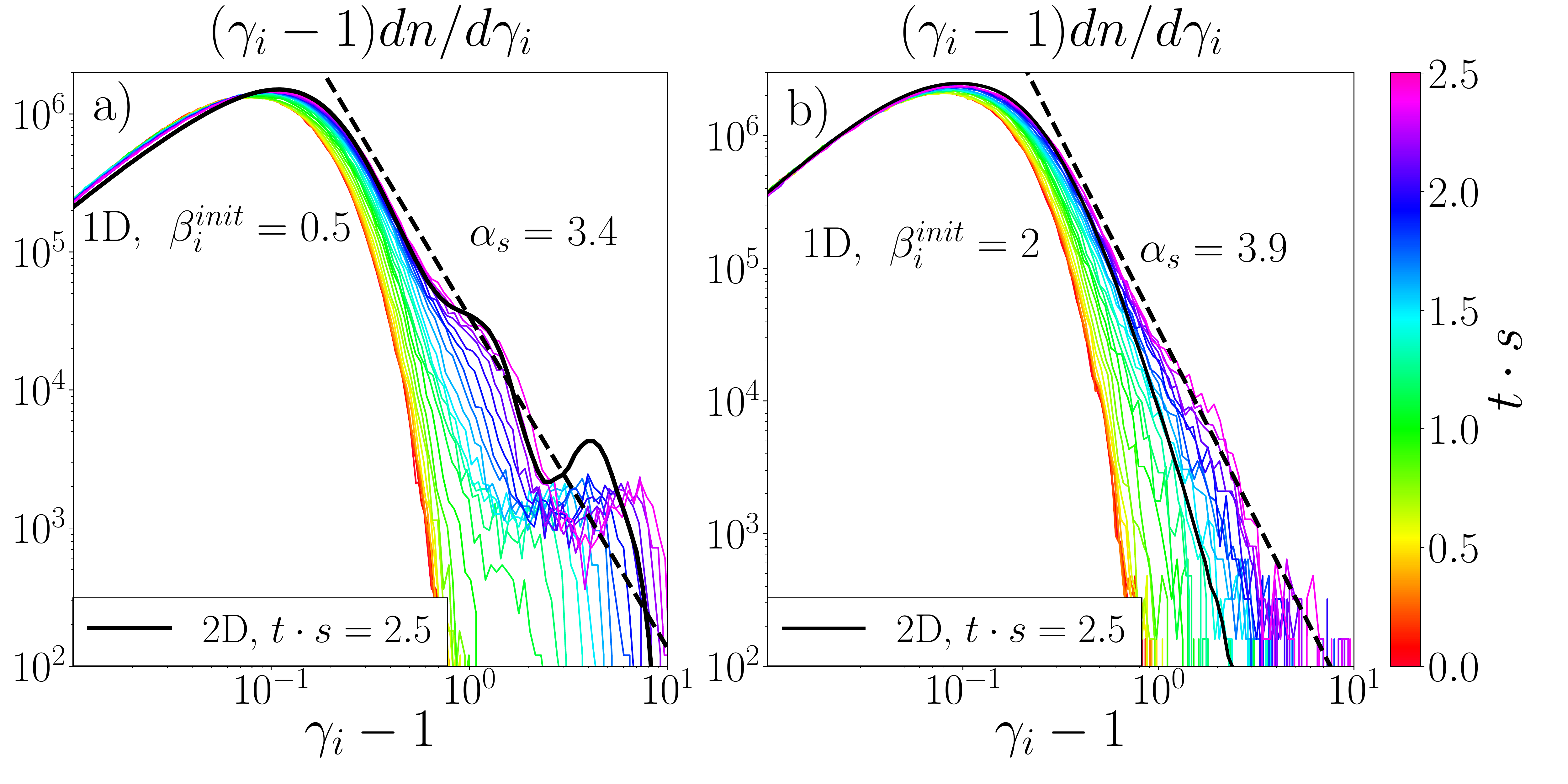}
\caption{\small Panels $a$ and $b$ show the evolution of the ion spectra for the 1D runs S1m2b0.5 ($m_i/m_e=2$, $\beta_{i}^{\textrm{init}}=0.5$) and S1m2b2 ($m_i/m_e=2$, $\beta_{i}^{\textrm{init}}=2$), respectively. The color bar indicates the time for each spectrum. The black lines show the final spectra ($t\cdot s=2.5$) of the 2D runs S2m2b0.5 ($m_i/m_e=2$, $\beta_{i}^{\textrm{init}}=0.5$; panel $a$) and S2m2b2 ($m_i/m_e=2$, $\beta_{i}^{\textrm{init}}=2$; panel $b$), which, apart from the different number of dimensions, assume the same plasma conditions as runs S1m2b0.5 and S1m2b2, respectively.} 
\label{1d2dspect} 
\end{figure}
\begin{figure*}[t!]  
\centering 
\vspace*{-0.2cm}
\hspace*{-0.3cm}
\includegraphics[width=18.2cm]{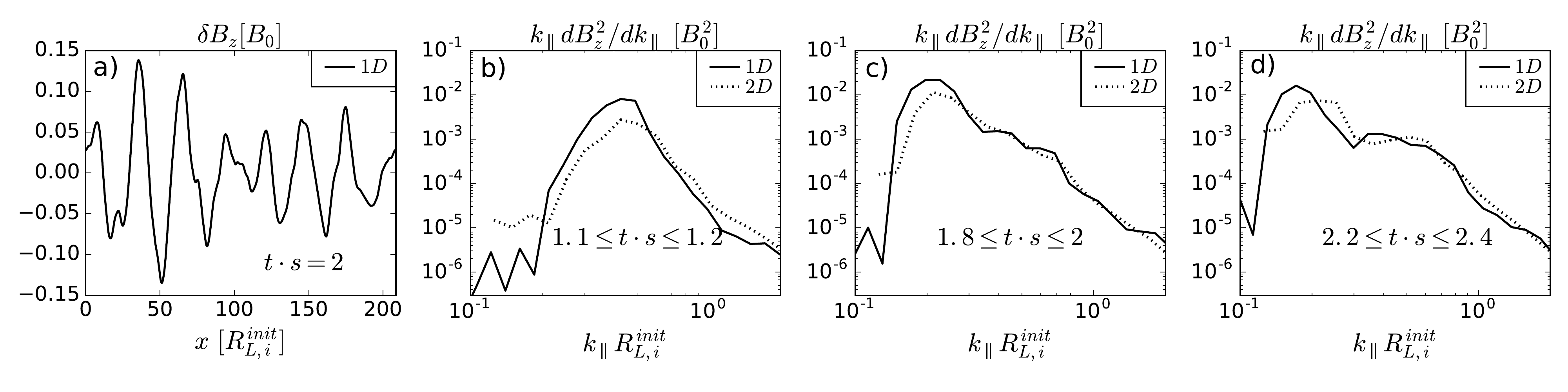}
\vspace*{-0.7cm}
\caption{Panel $a$ is a snapshot of $\delta B_z$ at $t\cdot s=2$ for the 1D run S1m2b0.5, which is analogous to the 2D run S2m2b0.5, whose $\delta B_z$ component at $t\cdot s=2$ is shown in Fig. \ref{fig:fields}$c$. Panels $b$, $c$, and $d$ show the average power spectra of $\delta B_z$, $k_{\parallel}dB_z^2/dk_{\parallel}$, for the same 1D and 2D runs during the periods $1.1< t\cdot s < 1.2$, $1.8< t\cdot s < 2$, and $2.2< t\cdot s < 2.4$, respectively. $k_{\parallel}$ denotes the $\textbf{\textit{k}}$ component parallel to $\textbf{\textit{B}}$ in the 2D case, which corresponds to simply $k$ in the 1D case.} 
\label{fig:powerfluct} 
\end{figure*}

\noindent Figure \ref{fig:spectra}$b$ shows the evolution of the ion spectra for 
the 2D run S2m2b2 ($m_i/m_e=2$ and $\beta_{i}^{\textrm{init}}=2$). In this case the growth of the non-thermal energy tail is also present but with a slower growth throughout the whole simulation. By $t\cdot s =2.5$ the tail can be approximated by a much less pronounced power law with spectral index $\alpha_s \approx 5.9$. The solid-black line represents the final spectrum for the analogous run S2m10b2, with $m_i/m_e=10$. The small difference between the $m_i/m_e=2$ and 10 cases suggests that, as in the IC-dominated case, the ion to electron mass ratio is fairly unimportant in determining the ion acceleration efficiency.\newline

\noindent These results strongly suggest that the presence of IC modes is key for the acceleration of ions. In \S \ref{sec:1D} we show that this is indeed the case making use of 1D simulations in which the mirror modes are artificially suppressed. In \S \ref{sec:1D} we will also make use of the low computational cost of 1D simulations to test the effect of using values of $m_i/m_e$ and $\omega_{c,i}^{\textrm{init}}/s$ much larger than the ones used in the 2D runs. As we will see, we will find no significant dependence of the ion acceleration on these parameters.

\section{1D shearing simulations} 
\label{sec:1D}
\noindent Since the IC modes propagate mainly parallel to the background magnetic field $\langle \textbf{\textit{B}} \rangle$, in this section we study the ion acceleration due to the IC instability by only capturing modes with wave vector $\textbf{\textit{k}}$ parallel to $\langle \textbf{\textit{B}} \rangle$. We do so by using the 1D version of our shear coordinates simulations \citep{RiquelmeEtAl2012}. Analogously to what happens in 2D and 3D, in our 1D shearing runs the spatial domain of the simulation evolves with time, following the shearing flow of the plasma. This implies that the domain rotates and stretches so that the resolved $\textbf{\textit{k}}$'s are always parallel to $\langle \textbf{\textit{B}} \rangle$. A detailed description of our 1D setup is in Appendix \ref{1druns}. \newline

\noindent In the next section, we show the suitability of the 1D setup to study problems dominated by modes with $\textbf{\textit{k}} \parallel \textbf{\textit{B}}$ by comparing 1D runs with 2D simulations that are dominated by the IC instability. Then, we use the 1D runs to: $i)$ provide further evidence that the IC modes are the essential ingredient for the ion acceleration, and $ii)$ explore the dependence of the ion acceleration on $m_i/m_e$ and $\omega_{c,i}^{\textrm{init}}/s$, which will make use of the low computational cost of the 1D runs.

\subsection{1D vs. 2D comparison}
\label{1d2dcompa}
\noindent We use runs S1m2b0.5 (1D) and S2m2b0.5 (2D) to compare the 1D and 2D results. First, we check whether our 1D and 2D runs give similar ion spectra. Fig. \ref{1d2dspect}$a$ shows the evolution of the ion spectrum for run S1m2b0.5 (1D) from $t\cdot s=0$ to $t\cdot s=2.5$. We see that the 1D spectral evolution is very similar to the one of the 2D run S2m2b0.5 shown in Fig \ref{fig:spectra}$a$. Indeed, both spectra can be described as a power law of index $\alpha_s\approx 3.4$ plus two bumps, with the intermediate energy bump appearing at $t\cdot s \gtrsim 2$. A more detailed comparison can be made by overplotting the final ($t\cdot s=2.5$) spectrum for run S2m2b0.5 in Figure Fig. \ref{1d2dspect}$a$ (black line). We see that the two final spectra are very similar, with the main difference being a $\sim 2$ times larger maximum energy in the 1D run. This small difference is to some extent expected, due to the presence of mirror modes in the 2D case. Indeed, in 2D we have some contributions of mirror modes, which however are not conducive to ion acceleration. This explains why the 2D setup leads to somewhat lower energy gains than in 1D.\newline

\noindent In terms of the ion pressure anisotropy, Fig. \ref{fig:cgl}$a$ shows $p_{i,\parallel}$ (red) and $p_{i,\perp}$ (blue) for runs S1m2b0.5 (1D; dotted) and S2m2b0.5 (2D; solid). Both for $p_{i,\parallel}$ and $p_{i,\perp}$, the 1D and 2D simulations give essentially the same results. In dashed lines we show the corresponding double-adiabatic behavior \citep{ChewEtAl1956}, which is followed quite well by the two  simulations until $t\cdot s \approx 1$. With respect to energy conservation, Figure \ref{fig:cgl}$b$ shows the volume average $dU_i/dt$ (black) and $r\Delta p_i$ (green) for the same runs S1m2b0.5 (1D; dashed) and S2m2b0.5 (2D; solid). $dU_i/dt$ behaves very similarly in 1D and 2D, and in the two cases it corresponds quite well to the heating due to anisotropic viscosity.\newline

\noindent In order to compare the behavior of the magnetic fluctuations $\delta \textbf{\textit{B}}$, we use $\delta B_z$ since in the 2D runs this component is mainly produced by the IC modes, and it is essentially not affected by the (subdominant but still present) mirror modes. Figure \ref{fig:powerfluct}$a$ shows a snapshot of $\delta B_z$ at $t\cdot s=2$ for the 1D run S1m2b0.5. We see that both in terms of the dominant wavelength ($\approx 30 R_{L,i}^{\textrm{init}}$) and of its amplitude, $\delta B_z$ behaves fairly similarly to the 2D case, shown in Fig. \ref{fig:fields}$c$. Figures \ref{fig:powerfluct}$b$, \ref{fig:powerfluct}$c$ and \ref{fig:powerfluct}$d$ show power spectra of $\delta B_z$ at different simulation times for a more detailed comparison. In order to reduce the effects of time variability, we take averages during $1.1< t\cdot s < 1.2$, $1.8< t\cdot s < 2$, and $2.2< t\cdot s < 2.4$, respectively. We see that the 1D and 2D spectra look quite similar. Their main difference consists of a small (by a factor $\sim 1.5$) shift in the peak of the 1D spectra towards longer wavelengths, and a factor $\sim 2$ increase in the peak amplitude.\newline 

\noindent These differences in the $\delta B_z$ spectra can be explained to a large extent by the small differences in the ion energy spectra. Indeed, in \S \ref{sec:resonance} we show that the wave number $k$ at the peak of the IC wave spectrum is determined by the resonance condition with the highest energy ions, with $k \propto 1/\gamma_i$. Thus, since the 1D runs produce maximum ion energies $\sim 2$ times larger than in the 2D case, the wave number at the peak should be reduced by a similar factor.\newline

\noindent Something similar occurs with the difference in amplitude of $\delta B_z$. For relativistic ions interacting resonantly with parallelly propagating waves, the effective scattering frequency $\nu_{eff,i}$ should scale as \citep{KulsrudEtAl1969}:
\begin{equation}
\nu_{eff,i} \propto \frac{\omega_{c,i}}{B_0^2\gamma_i}\frac{d(\delta B_z^2)}{d\ln(k)}.
\label{nu}
\end{equation}        
 
\noindent Thus, since the ions in the 1D and 2D runs are scattered at roughly the same rate (given their similar evolution of $p_{\parallel,i}$ and $p_{\perp,i}$, as shown in Fig. \ref{fig:cgl}$a$), Eq. \ref{nu} implies that the peak value of $dB_z^2{k}/dk$ in the 1D case should roughly be $\sim 2$ times larger than in the 2D case, which is seen in panels $b$, $c$ and $d$ of Fig. \ref{fig:powerfluct}.\newline

\noindent Thus, besides a factor $\sim 2$ difference in the highest energy of ions (which is likely due to the weak presence of mirror modes in the 2D runs), the 1D runs reproduce reasonably well the 2D results, and provide a valuable tool to study the effect of IC waves on ion acceleration. In the next two sections we use 1D runs to provide further evidence that the IC modes are indeed the essential ingredient for the ion acceleration, and to explore the dependence of the acceleration on $m_i/m_e$ and $\omega_{c,i}^{\textrm{init}}/s$.

\subsection{The role of IC and mirror modes}
\label{proof}
\noindent Simulations in 1D can be used to further clarify the role of IC and mirror modes in the acceleration of ions. We do this by comparing the ion spectra from a 2D simulation where the mirror modes dominate ($\beta_i^{\textrm{init}}=2$) with an analogous 1D run where these modes are artificially suppressed. This is done in Fig. \ref{1d2dspect}$b$, which shows in black the final spectrum of the 2D run S2m2b2, where the mirror modes dominate. In addition, Fig. \ref{1d2dspect}$b$ shows the ion spectra at different times for the 1D run S1m2b2, where the plasma is under the same conditions as in run S2m2b2. It can be seen that ions tend to be significantly more accelerated in the 1D case, in which the IC modes dominate. This shows that the main ingredient for the acceleration of ions is indeed the scattering by the IC modes, with the mirror instability suppressing the acceleration. This suppression is in line with the lack of electric field associated with the mirror modes, as shown in Fig. \ref{fig:fields}. Thus, when the mirror modes dominate, the scattering of ions tends to be elastic, and no acceleration effect should be present.

\subsection{$m_i/m_e$ and $\omega_{c,i}^{\textrm{init}}$ dependence}
\label{mimewcisdepend}
\begin{figure}[t!]
\vspace*{-0.2cm}
\subfloat{
  \centering
  \hspace*{-0.4cm}
  \includegraphics[width=0.26\textwidth]{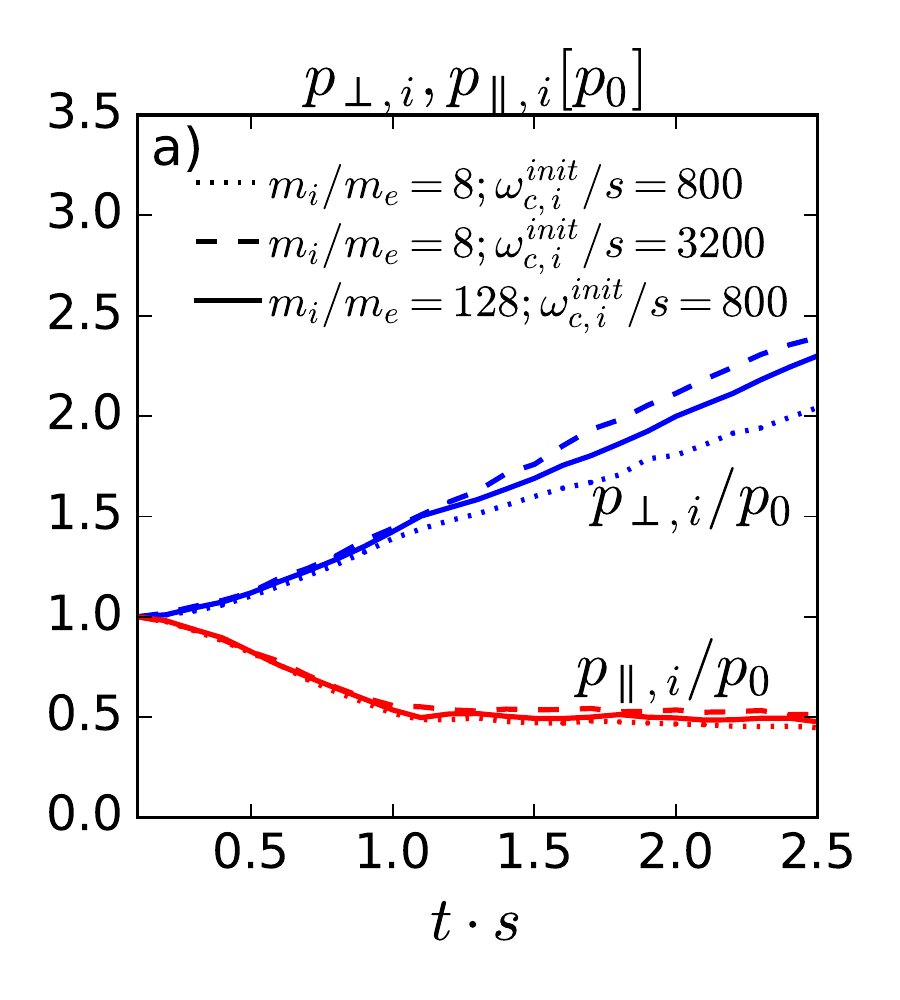}
}
\subfloat{
  \centering
  \hspace*{-0.55cm}
  \includegraphics[width=0.26\textwidth]{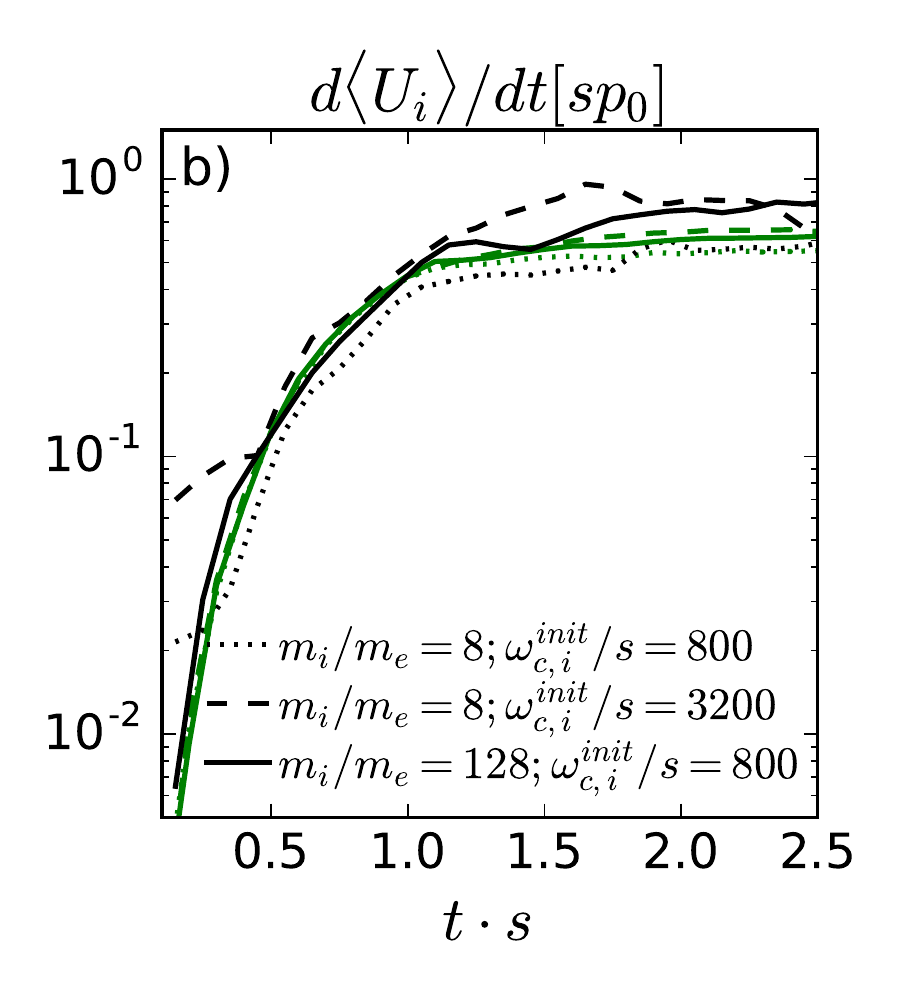}
}
\vspace*{-0.4cm}
\caption{\small Panel $a$: $p_{i,\parallel}$ (red) and $p_{i,\perp}$ (blue) for 1D runs S1m8b0.5 (dotted), S1m128b0.5 (solid) and S1m8b0.5d (dashed), all with $\beta_i^{\textrm{init}}=0.5$. Panel $b$: the volume-averaged $dU_i/dt$ (black) and $r\Delta p_i$ (green) for the same 1D runs and using the same line styles. Both $dU_i/dt$ and $r\Delta p_i$ evolve very similarly in all the runs, and in all cases the ion energy gain reproduces reasonably well the ``anisotropic viscosity" prediction $r\Delta p_i$ (Equation \ref{anvis}).}
\label{mimewcispres}
\end{figure}
\begin{figure}[t!]
\vspace*{-0.15cm}
\subfloat{
  \centering
  \hspace*{-0.4cm} 
  \includegraphics[width=0.26\textwidth]{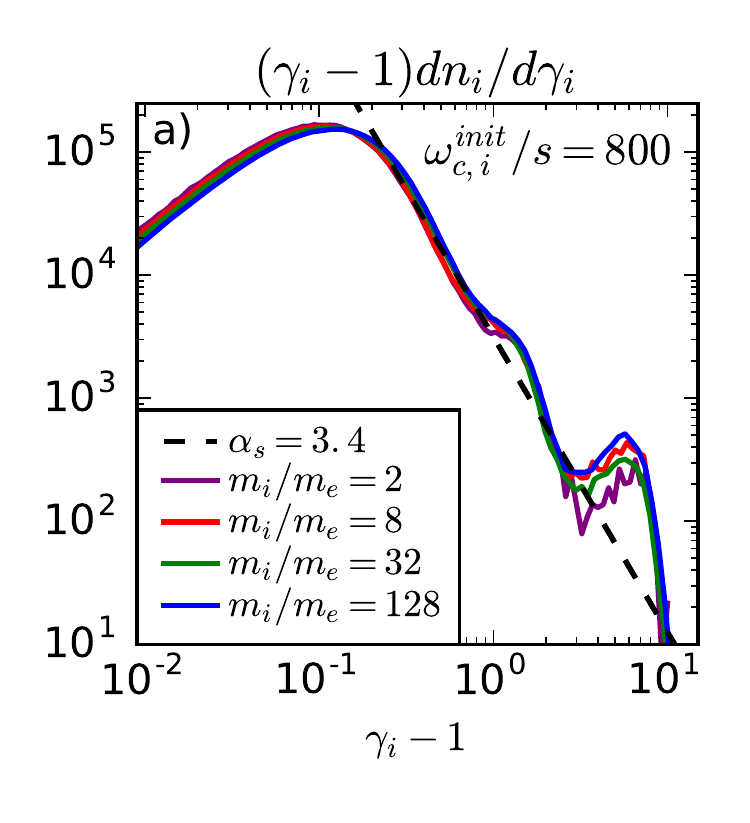}
}
\subfloat{
  \centering
  \hspace*{-0.55cm}
  \includegraphics[width=0.26\textwidth]{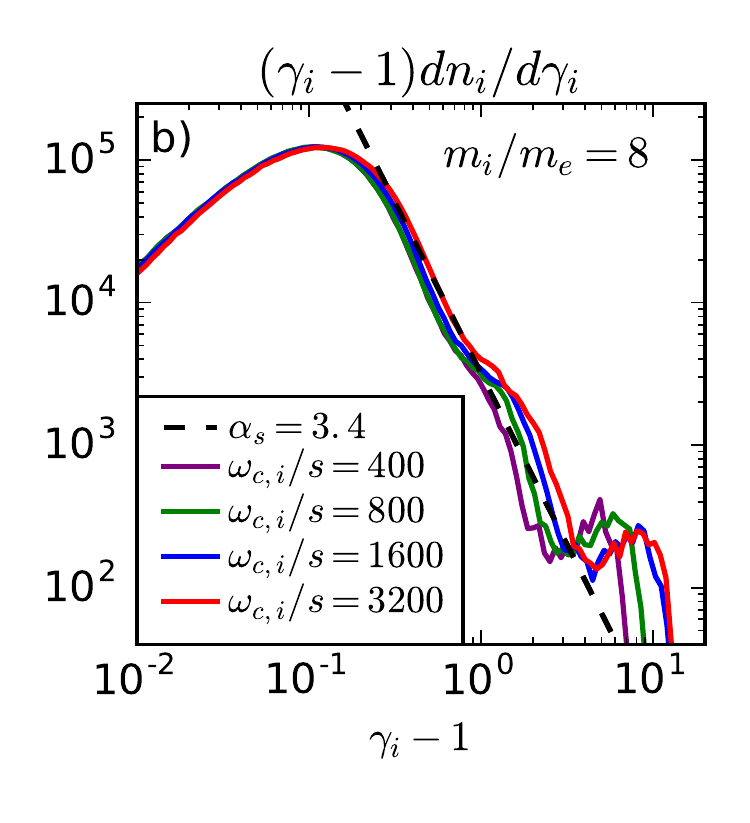}
}
\vspace*{-0.4cm}
\caption{\small The final ion spectra ($t\cdot s=2.5$) for 1D simulations with the same parameters $\beta_i^{\textrm{init}}=0.5$ and $k_BT_i=k_BT_e=0.05m_ic^2$. The runs in panel $a$ have $\omega_{c,i}^{\textrm{init}}=800$ and $m_i/m_e=2$, 8, 32, and 128 (respectively, runs S1m2b0.5, S1m8b0.5, S1m32b0.5, and S1m128b0.5 in Table \ref{table}). The runs in panel $b$ have $m_i/m_e=8$ but with $\omega_{c,i}^{\textrm{init}}/s=400$, 800, 1600 and 3200 (runs S1m8b0.5b, S1m8b0.5, S1m8b0.5c, and S1m8b0.5d, respectively).}
\label{mimewcis}
\end{figure}
\noindent We use 1D simulations to explore the dependence of the ion acceleration by the IC instability on $m_i/m_e$ and $\omega_{c,i}^{\textrm{init}}/s$, focusing on the case with $\beta_i^{\textrm{init}}=0.5$. In terms of the evolutions of $p_{\perp,i}$ and $p_{\parallel,i}$, Figure \ref{mimewcispres}$a$ shows the cases of runs S1m8b0.5 ($\omega_{c,i}^{\textrm{init}}/s=800$, $m_i/m_e=8$; dotted line), S1m128b0.5 ($\omega_{c,i}^{\textrm{init}}/s=800$, $m_i/m_e=128$; solid line) and S1m8b0.5d ($\omega_{c,i}^{\textrm{init}}/s=3200$, $m_i/m_e=8$; dashed line). No significant difference can be seen between the different mass ratios and magnetizations. The same thing happens when we look at the ion energy gain. Figure \ref{mimewcispres}$b$ shows $dU_i/dt$ (black) and $r\Delta p_i$ (green) for the same runs. We see that the ion energy gain is fairly independent of $m_i/m_e$ and $\omega_{c,i}^{\textrm{init}}/s$, and in all cases it reasonably well agrees with the heating prediction through anisotropic viscosity.\newline 

\noindent Figure \ref{mimewcis}$a$ shows the final ion spectra ($t\cdot s=2.5$) for simulations with $m_i/m_e=2$, 8, 32, and 128 (runs S1m2b0.5, S1m8b0.5, S1m32b0.5, and S1m128b0.5 in Table \ref{table}). In all simulations the ions share the same parameters: $\omega_{c,i}^{\textrm{init}}/s=800$, $\beta_i^{\textrm{init}}=0.5$, and $k_BT_i=k_BT_e=0.05m_ic^2$, so the only difference is the value of $m_i/m_e$. The non-thermal ion tail in all cases can be fairly well described as a power-law of spectral index $\alpha_s \approx 3.4$ plus two bumps occurring at roughly the same energies. This result shows that, as long as the electrons are somewhat less massive than the ions, their effect on the ion acceleration by IC modes becomes negligible. This is expected given the resonant nature of the interaction between the unstable IC modes and the ions, which requires the modes to be left-handed, circularly polarized \citep[e.g.,][]{Gary1992}. This polarization requirement naturally makes it significantly more difficult for the electrons to interact resonantly with the IC modes, even for a mass ratio as small as $m_i/m_e=2$. \newline
\begin{figure*}[t!]  
\centering 
\vspace*{-0.4cm}
\includegraphics[width=15cm]{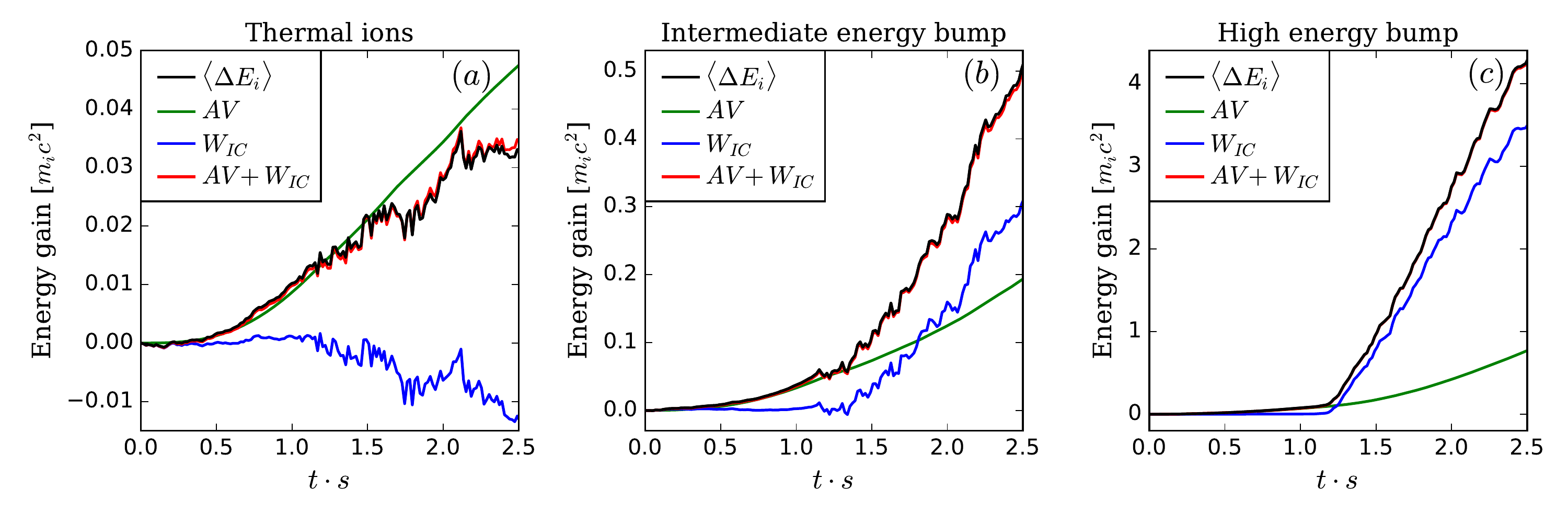}
\vspace*{-0.4cm}
\caption{Panels $a$, $b$ and $c$ show, for three different ion populations, the energy gain due to the work done by the electric field of the IC waves, $W_{IC}$ (blue line), and due to the ion anisotropic viscosity, $AV$ (green line). All the energies are normalized by $m_ic^2$. Panels $a$, $b$ and $c$ correspond to the ``thermal", ``intermediate energy" and ``highest energy" ions, which are chosen so that at $t\cdot s=2.5$ they have $0.09 < \gamma_i -1 < 0.1$, $1 < \gamma_i -1 < 1.03$ and $4.8 < \gamma_i -1$. In the three cases, we show that the sum of $W_{IC}$ and $AV$ (in red) corresponds quite well to the average variation in the ion energy, $\langle \Delta E_i \rangle$ (in black).} 
\label{fig:testp} 
\end{figure*}

\noindent Similarly, Figure \ref{mimewcis}$b$ shows the final spectra for simulations with $m_i/m_e=8$, $\beta_i^{\textrm{init}}=0.5$, and $k_BT_i=k_BT_e=0.05m_ic^2$, but with $\omega_{c,i}^{\textrm{init}}/s=400$, 800, 1600 and 3200 (runs S1m8b0.5b, S1m8b0.5, S1m8b0.5c and S1m8b0.5d, respectively). The spectra get slightly harder as $\omega_{c,i}^{\textrm{init}}/s$ increases, with the difference between them being progressively less significant as $\omega_{c,i}^{\textrm{init}}/s$ grows. However, in all cases the tail can be well described as a power-law of spectral index $\alpha_s \approx 3.4$ plus two bumps. \newline

\noindent The independence of the ion acceleration on $\omega_{c,i}^{\textrm{init}}/s$ can be inferred from the way the effective ion scattering rate $\nu_{eff,i}$ provided by the IC waves is related to $s$. This scattering rate can be estimated from the evolution of $p_{\parallel,i}$ in a homogeneous plasma with no heat flux, assuming that $B$ evolves on time and length scales much larger than $\omega_{c,i}^{-1}$ and $R_{L,i}$ (which is the case in our runs). This evolution is given by Eq. 1 of \cite{SharmaEtAl2007}:
\begin{equation}
\frac{\partial p_{\parallel,i}}{\partial t} + \nabla \cdot (p_{\parallel,i}\textbf{\textit{v}}) + 2p_{\parallel,i}\hat{\textbf{\textit{b}}}\hat{\textbf{\textit{b}}}:\nabla \textbf{\textit{v}}=\frac{2}{3}\nu_{eff,i}\Delta p,
\label{prateek1}
\end{equation}
where $\textbf{\textit{v}}$ is the plasma bulk velocity and $\hat{\textbf{\textit{b}}}\equiv \textbf{\textit{B}}/B$. In the case of the shearing plasma ($\textbf{\textit{v}}=-sx\hat{y}$), $\nabla \cdot (p_{\parallel,i}\textbf{\textit{v}})=0$ and $\hat{\textbf{\textit{b}}}\hat{\textbf{\textit{b}}}:\nabla \textbf{\textit{v}}=\hat{b}_x\hat{b}_ys$. Fig. \ref{mimewcispres}$a$ shows that, after the saturation of the IC modes, $p_{\parallel,i}$ changes at a rate much smaller than $s$ so we can approximate $\partial p_{\parallel,i}/\partial t \approx 0$. Additionally, $\hat{b}_x\hat{b}_y$ ranges between 0.5 at $t\cdot s=1$ and 0.34 at $t\cdot s=2.5$. Thus, we simply assume $\hat{b}_x\hat{b}_y\approx 1/2$ and Eq. \ref{prateek1} becomes:
\begin{equation}  
\nu_{eff,i} \approx \frac{3}{2}s (p_{\parallel,i}/\Delta p_i).
\label{nus}
\end{equation}
By comparing the evolutions of $p_{\perp,i}$ and $p_{\parallel,i}$ for runs S1m8b0.8 ($\omega_{c,i}^{\textrm{init}}/s=800$; dot-dashed line) and S1m8b0.8d ($\omega_{c,i}^{\textrm{init}}/s=3200$; dashed line), Figure \ref{mimewcispres}$a$ shows that the factor $p_{\parallel,i}/\Delta p_i$ is fairly independent of $\omega_{c,i}^{\textrm{init}}/s$. Thus, Eq. \ref{nus} implies that $\nu_{eff,i} \propto s$.\newline

\noindent This proportionality between $\nu_{eff,i}$ and $s$ means that the average number of scatterings experienced by the ions after $t=2.5s^{-1}$ (at the end of the simulations) should be about the same in all runs. This property, if the IC modes properties are the same in all the simulations (as it occurs with the runs shown in Fig. \ref{mimewcis}$b$), should make the accelerating effect of the IC modes independent of $\omega_{c,i}^{\textrm{init}}/s$.\newline

\noindent Our 1D simulations, therefore, show that the ion acceleration by IC modes is fairly independent of $m_i/m_e$ and $\omega_{c,i}^{\textrm{init}}/s$. In the next section, we use 1D and 2D simulations to describe in further detail the way that ions of different energy get accelerated, emphasizing the role played by their resonant interaction with the IC waves.

\section{The acceleration mechanism} 
\label{sec:nature}
\noindent The two possible sources of energy for the ions in our simulations are: the energy gain due to anisotropic viscosity (Eq. \ref{anvis}), and the energy gain due to the work done by the electric field associated with the IC waves (considering that the electric field associated to the mirror modes is negligible). In \S \ref{sec:stochasticvsvisco} we identify the contributions of each of these energy sources to producing the non-thermal ion spectra.

\subsection{IC work vs. anisotropic viscosity} 
\label{sec:stochasticvsvisco}
\noindent Figure \ref{fig:testp} shows the different contributions to the energy gain of three ion populations from the 2D run S2m2b0.5, separated according to their final energy at $t\cdot s=2.5$. These populations are:
\begin{enumerate}
\item The ``thermal ions", corresponding to ions in the bulk of the ion distribution, and with their energy gain plotted in Fig. \ref{fig:testp}$a$. These ions are chosen so that at $t\cdot s=2.5$ their Lorentz factors satisfy $0.09 < \gamma_i-1 < 0.1$ (marked by the vertical red line in Fig. \ref{fig:spectra}$a$).   
\item The ions in the ``intermediate energy" bump of the tail (shown in Fig. \ref{fig:testp}$b$). These ions have Lorentz factors in the range $1 < \gamma_i-1 < 1.03$ at $t\cdot s=2.5$ (marked by the vertical dark green line in Fig. \ref{fig:spectra}$a$).
\item The ions in the ``high energy" bump of the tail (shown in Fig. \ref{fig:testp}$c$). This population corresponds to the highest energy ions, defined by $\gamma_i-1 > 4.8$ at $t\cdot s=2.5$ (marked by the grey region in Fig. \ref{fig:spectra}$a$).
\end{enumerate}
For each of these populations we plot the following contributions to their energy gain: 
\begin{enumerate}
\item The work done by the electric field of the IC waves, $W_{IC}$, which is shown by the blue lines of Fig. \ref{fig:testp}.
\item The energy gain by anisotropic viscosity, $AV$, shown by the green line in Fig. \ref{fig:testp}.\footnote{For each ion population, this energy gain is calculated as the integral in time of the rate of energy gain due to viscosity: $\int dt \,r\,\Delta p_i$ (see Eq. \ref{anvis}), where $\Delta p_i$ is calculated using only the ions of each population.} 
\end{enumerate}

\noindent The blue line in Figure \ref{fig:testp}$a$ shows that the energy given by the electric field of the IC waves to the thermal ions, $W_{IC}$, is negative. This implies that the scattering process on average substracts energy from the thermal ions and transfers it to the waves. The total gain in energy of the thermal ions is still positive, and dominated by viscous heating. On the other hand, Figure \ref{fig:testp}$b$ shows that the work done by the IC waves on the ions of the intermediate energy bump is positive and larger than the heating by anisotropic viscosity, $AV$, which means that these ions are mainly energized by the scattering caused by the IC waves. This energization occurs mainly after $t\cdot s \approx 2$, which is consistent with the late time appearance of the intermediate energy bump, as shown by the time evolution of the ion spectrum depicted in Figure \ref{fig:spectra}$a$. Finally, Figure \ref{fig:testp}$c$ shows that $W_{IC}$ is about three times larger than $AV$, implying that the IC acceleration for the highest energy ions is even larger than for the ``intermediate energy" ions.\newline

\noindent In the three Figures \ref{fig:testp}$a$, \ref{fig:testp}$b$ and \ref{fig:testp}$c$ we also plot the sum of $AV$ and $W_{IC}$ (red line) and the average change in energy of the three ion populations, $\langle \Delta E_i \rangle$ (black line). We see that these two quantities are essentially the same for the three populations, implying that the energy gain due to anisotropic viscosity and the electric field of the IC modes accounts quite well for the total ion energy evolution in the three populations.   

\subsection{Resonance with IC waves} 
\label{sec:resonance}

In \S \ref{sec:stochasticvsvisco} we show that a non-thermal ion tail is produced by the scattering of ions off IC waves, which, in turn, obtain their energy from the pressure anisotropy of the thermal ions. This implies that this acceleration mechanism requires the resonance condition between ions and IC waves to be satisfied by both the thermal and non-thermal ions. The resonance condition is:
\begin{equation}
\frac{\omega}{k} - v_{||} = \frac{\omega_{c,i}}{\gamma_ik},
\label{eq:res}
\end{equation}
where $\omega$ and $k$ are the real part of the frequency and the wave number of the modes, $v_{||}$ is the ion velocity parallel to \textbf{\textit{B}}, and $\omega_{c,i}$ ($\equiv eB/m_ic$) is the non-relativistic cyclotron frequency of the ions. \newline
\begin{figure}
  \centering
 \hspace*{-0.27cm} 
 \vspace*{-0.4cm}
 \includegraphics[width=.5\textwidth]{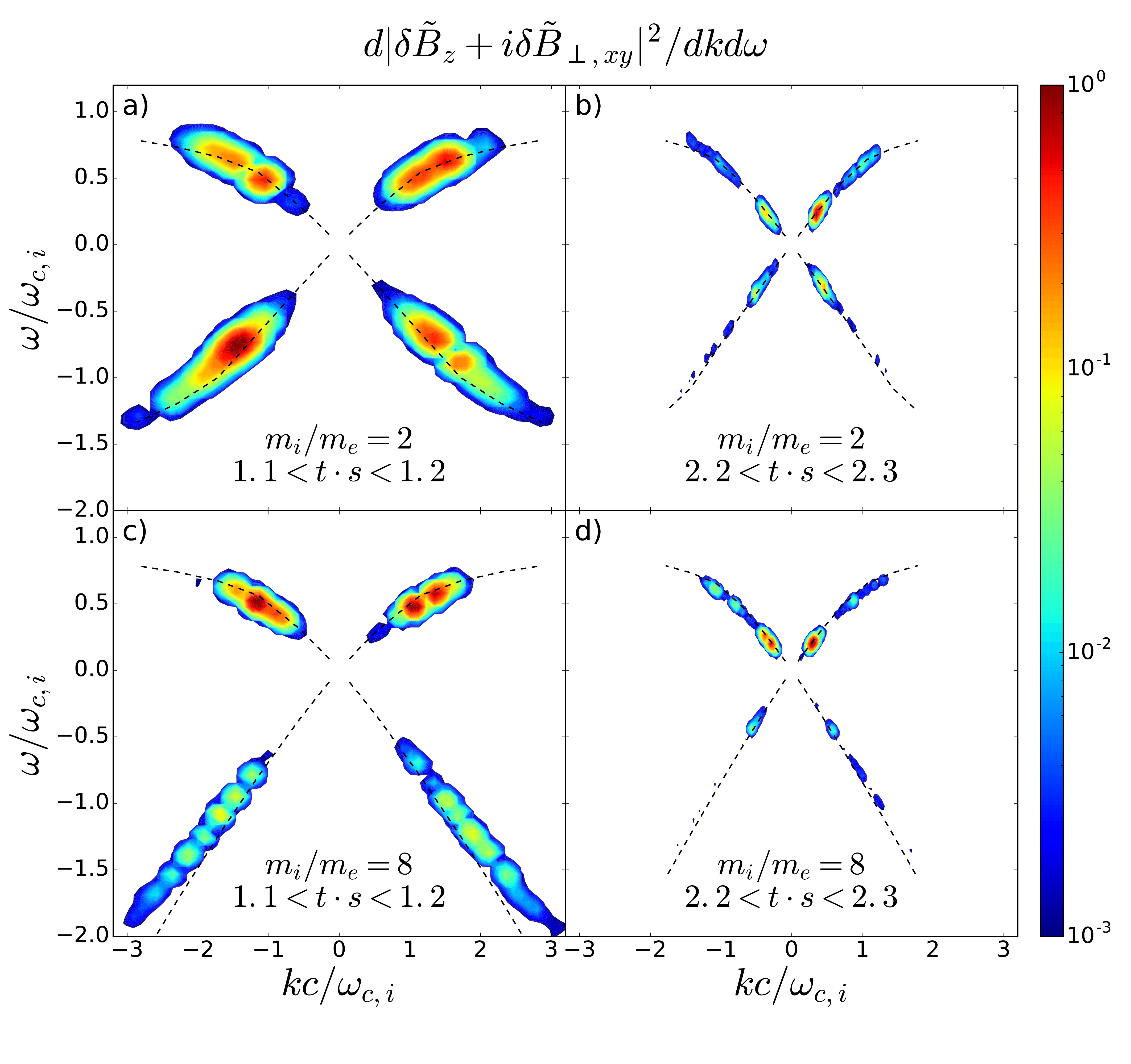}
 \vspace*{-0.4cm}
\caption{\small The four panels show $|\delta \tilde{B}_z(\omega,k) + i \delta \tilde{B}_{\perp, xy}(\omega,k)|^2$ for different 1D simulations and time intervals, where $\delta \tilde{B}_z(\omega,k)$ and $\delta \tilde{B}_{\perp, xy}(\omega,k)$ are the Fourier transforms in time and space of the two mutually perpendicular components of $\delta \textbf{\textit{B}}$, $\delta B_z$ and $\delta B_{\perp, xy}$ (both perpendicular to $\langle \textbf{\textit{B}} \rangle$). The combination $\delta \tilde{B}_z(\omega,k) + i \delta \tilde{B}_{\perp, xy}(\omega,k)$ makes the contributions to $\delta \textbf{\textit{B}}$ of the circularly polarized IC waves (left-handed) and whistler waves (right-handed) appear at $\omega > 0$ and $< 0$, respectively. Panels $a$ and $b$ correspond to run S1m2b0.5 ($\beta_i^{\textrm{init}}=0.5$, $m_i/m_e=2$, $\omega_{c,i}^{\textrm{init}}/s=800$) at $1.1 < t\cdot s < 1.2$ and $2.2 < t\cdot s < 2.3$, respectively. Panels $c$ and $d$ correspond to run S1m8b0.5 ($\beta_i^{\textrm{init}}=0.5$, $m_i/m_e=8$, $\omega_{c,i}^{\textrm{init}}/s=800$) also at $1.1 < t\cdot s < 1.2$ and $2.2 < t\cdot s < 2.3$, respectively. $\omega$ is normalized to the ``instantaneous" cyclotron frequency $\omega_{c,i}$ ($\equiv \omega_{c,i}^{\textrm{init}}B/B_0$), while $k$ is normalized to $\omega_{c,i}/c$. In dashed lines, we show the theoretical IC and whistler dispersion relations $\omega_{\textrm{theo}}(k)$ obtained with the linear Vlasov solver NHDS \citep{VerscharenEtAl2018}.}
\label{fig:disprel}
\end{figure}

\noindent In order to check that this resonance condition is satisfied by ions of all energies, in Figs. \ref{fig:disprel}$a$ and \ref{fig:disprel}$b$ we measure the ratio $\omega/k$ for the IC waves at two different time intervals for the 1D run S1m2b0.5 ($\beta_i^{\textrm{init}}=0.5$, $m_i/m_e=2$, $\omega_{c,i}^{\textrm{init}}=800$). We do this by plotting $|\delta \tilde{B}_z(\omega,k) + i \delta \tilde{B}_{\perp, xy}(\omega,k)|^2$, where $\delta B_z(\omega,k)$ and $\delta B_{\perp, xy}$ correspond to two mutually perpendicular components of $\delta \textbf{\textit{B}}$, that are also perpendicular to $\langle \textbf{\textit{B}} \rangle$ (see caption of Fig. \ref{fig:evolb}), the tilde ($\,\tilde{ }\,$) denotes the time and space Fourier transform of a quantity and $i=\sqrt{-1}$. Thus, the combination $\delta \tilde{B}_z(\omega,k) + i \delta \tilde{B}_{\perp, xy}(\omega,k)$ allows to separate the contributions to $\delta \textbf{\textit{B}}$ provided by IC waves (left-handed, circularly polarized) and whistler waves (right-handed, circularly polarized), with the latter being expected to be destabilized by the pressure anisotropy of electrons \citep{GaryEtAl1996}. In the case of run S1m2b0.5, the IC waves contribute to $|\delta \tilde{B}_z(\omega,k) + i \delta \tilde{B}_{\perp, xy}(\omega,k)|$ only for $\omega > 0$, while the whistler wave contribution appears for $\omega < 0$. This way, calculating $|\delta \tilde{B}_z(\omega,k) + i \delta \tilde{B}_{\perp, xy}(\omega,k)|^2$ allows to separate the IC and whistler contributions to $\delta \textbf{\textit{B}}$, and to estimate $\omega(k)$ for these two modes.\newline

\noindent Figure \ref{fig:disprel}$a$ corresponds to the time interval $1.1 < t\cdot s < 1.2$ of run S1m2b0.5. The IC modes have a phase velocity of $\omega/k \approx 0.5c$. Additionally, we measure the rms ion velocity parallel to $\textbf{\textit{B}}$ at $1.1 < t\cdot s < 1.2$, which is $v_{\parallel}^{\textrm{rms}}\approx 0.16c$,\footnote{This can be estimated by the factor $\sim 2$ decrease in $p_{i,\parallel}$ seen in Fig. \ref{fig:cgl}$a$, and considering that initially $k_BT_i/m_ic^2=0.05$.} implying that to a good approximation we can neglect the $v_{\parallel}$ term on the left hand side of Eq. \ref{eq:res}. The resonance condition at $1.1 < t\cdot s < 1.2$ can thus be written as:

\begin{equation}
\gamma_i(kc/\omega_{c,i}) \approx 2.
\end{equation} 
\noindent At $1.1 < t\cdot s < 1.2$, $\gamma_i$ is in the range $1 < \gamma_i\lesssim 2$ (see Fig. \ref{1d2dspect}$a$), implying that most of the power of the IC modes should be in the range $1\lesssim kc/\omega_{c,i}\lesssim 2$, which coincides well with the range of $k$ in which most of the IC power is observed in Fig. \ref{fig:disprel}$a$. \newline

\noindent Analogously, Fig. \ref{fig:disprel}$b$ shows $|\delta B_z(\omega,k) + i \delta B_{\perp, xy}(\omega,k)|^2$ for the same simulation but in the time range $2.2 < t\cdot s < 2.3$. In this case, Fig. \ref{fig:disprel}$b$ shows that $\omega/k \approx 0.7c$. Thus, making a similar analysis as in the case $1.1 < t\cdot s < 1.2$, we obtain:\footnote{Here we also assume $\omega/k$ ($\approx 0.7c$) $ \gg v_{\parallel}$, which allows neglecting $v_{\parallel}$ in Eq. \ref{eq:res}. This is a reasonable approximation considering that at $2.2 < t\cdot s < 2.3$, $v_{\parallel}$ (measured directly from the simulation) is always smaller than $\sim 0.4c$, even considering the highest energy particles in the non-thermal tail.}
\begin{equation}
\gamma_i(k c/\omega_{c,i}) \approx 1.5.
\label{eq3}
\end{equation}
Since in this time interval $1 < \gamma_i\lesssim 8$, we obtain that most of the power of the IC modes should be in the range $0.2 \lesssim kc/\omega_{c,i} \lesssim 1.5$. This interval coincides reasonably well with the range of $k$ where most power is concentrated in Fig. \ref{fig:disprel}$b$. Notice that this power appears to be enhanced in two intervals of $k$. The high-$k$ interval corresponds to $0.7\lesssim kc/\omega_{c,i} \lesssim 1.5$, which, according to Eq. \ref{eq3}, implies resonance with ions with $1 < \gamma_i\lesssim 2$. Remarkably, this is a range of $\gamma_i$ with abundant IC scattering at $2.2 \lesssim t\cdot s \lesssim 2.3$, as shown by the rapid formation of the ``intermediate energy" bump, which mainly occurs at $2 \lesssim t\cdot s \lesssim 2.5$. The low-$k$ enhancement occurs for $0.2\lesssim kc/\omega_{c,i} \lesssim 0.5$, which, according to Eq. \ref{eq3}, corresponds to $3 \lesssim \gamma_i \lesssim 8$. This $\gamma_i$ interval coincides well with the ``high energy" bump shown in Fig. \ref{1d2dspect}$a$ at $t\cdot s \sim 2.2-2.3$, and is also consistent with the rapid increase in energy of this bump.\newline

\noindent Thus, we have shown that the range of $k$ in which the amplitude of the IC modes is significant is consistent with the resonance condition occurring for both the thermal and non-thermal ions in the tail. Figs \ref{fig:disprel}$c$ and \ref{fig:disprel}$d$ show the same quantities as Figs \ref{fig:disprel}$a$ and \ref{fig:disprel}$b$ but for simulation S1m8b0.5, where the ions are under the same conditions as in run S1m2b0.5, but with $m_i/m_e=8$ instead of $m_i/m_e=2$. The quantity $|\delta \tilde{B}_z(\omega,k) + i \delta \tilde{B}_{\perp, xy}(\omega,k)|^2$ essentially preserves the same properties for the IC modes ($\omega > 0$). The fact that $\omega(k)$ of the IC modes is fairly independent of $m_i/m_e$ is consistent with the near independence of the ion acceleration process on $m_i/m_e$. The whistler modes, on the other hand, do change significantly their properties as $m_i/m_e$ is increased, for instance by getting weaker and increasing their frequency. This is expected since quantities like $\omega_{c,e}/\omega_{p,e}$, $\omega_{c,e}/s$ and $k_BT_e/m_ec^2$ do change when varying $m_i/m_e$.\newline

\noindent Finally, the four panels in Fig. \ref{fig:disprel} also include IC and whistler theoretical dispersion relation calculations, $\omega_{\textrm{theo}}(k)$ (dashed lines), obtained with the linear Vlasov solver NHDS \citep{VerscharenEtAl2018}. These calculations assume the ion and electron conditions obtained in the simulations in terms of their temperatures, pressure anisotropies, mass ratio and Alfv\'en velocity. However, they do not consider departures from Maxwell-Boltzmann distributions or relativistic effects. In the four panels $\omega_{\textrm{theo}}(k)$ reproduces the behavior of $|\delta \tilde{B}_z(\omega,k) + i \delta \tilde{B}_{\perp, xy}(\omega,k)|^2$ for the case of the IC modes well, which shows that the phase (and group) velocity of the IC waves obtained from the simulations are not significantly affected by the non-thermal or relativistic effects in the ion velocity distribution. Most discrepancies occur for whistler waves when $m_i/m_e=8$. In this case, the discrepancy is most likely due to the electrons being significantly relativistic (Lorentz factor $\gamma_e \gg 1$), which is a regime strictly not captured by NHDS.\newline

\section{Comparing shear vs. compression} 
\label{sec:comp}

\noindent The ion acceleration presented in this paper occurs during the non-linear, saturated stage of the IC instability. This stage is reached through the continuous amplification of a background magnetic field $\langle\textbf{\textit{B}}\rangle$, which in previous sections has been driven by an imposed shear plasma motion. In this section we show that, as long as the ion conditions are similar, the specific process that amplifies $\langle\textbf{\textit{B}}\rangle$ does not play an important role in the acceleration. We do this by running 1D simulations in which $\langle\textbf{\textit{B}}\rangle$ grows due to plasma compression. The simulation setup is the same as in \cite{SironiEtAl2015}. The plasma is compressed along the $y$ and $z$ axes, with $\langle\textbf{\textit{B}}\rangle$ pointing along $x$. In this setup, $\langle\textbf{\textit{B}}\rangle$ is amplified due to magnetic flux conservation \citep[see Fig. 1 of][]{SironiEtAl2015}. In this situation, the background field grows as $|\langle\textbf{\textit{B}}\rangle|=B_0 (1+qt)^2$, where $q$ is a constant that roughly corresponds to the compression rate of the plasma. These 1D simulations only capture modes with $\textbf{\textit{k}} \parallel \langle\textbf{\textit{B}}\rangle$, which means that the mirror modes are artificially suppressed.\newline
\begin{figure}
\subfloat{
  \centering
\hspace*{-0.7cm}
\includegraphics[width=0.515\textwidth]{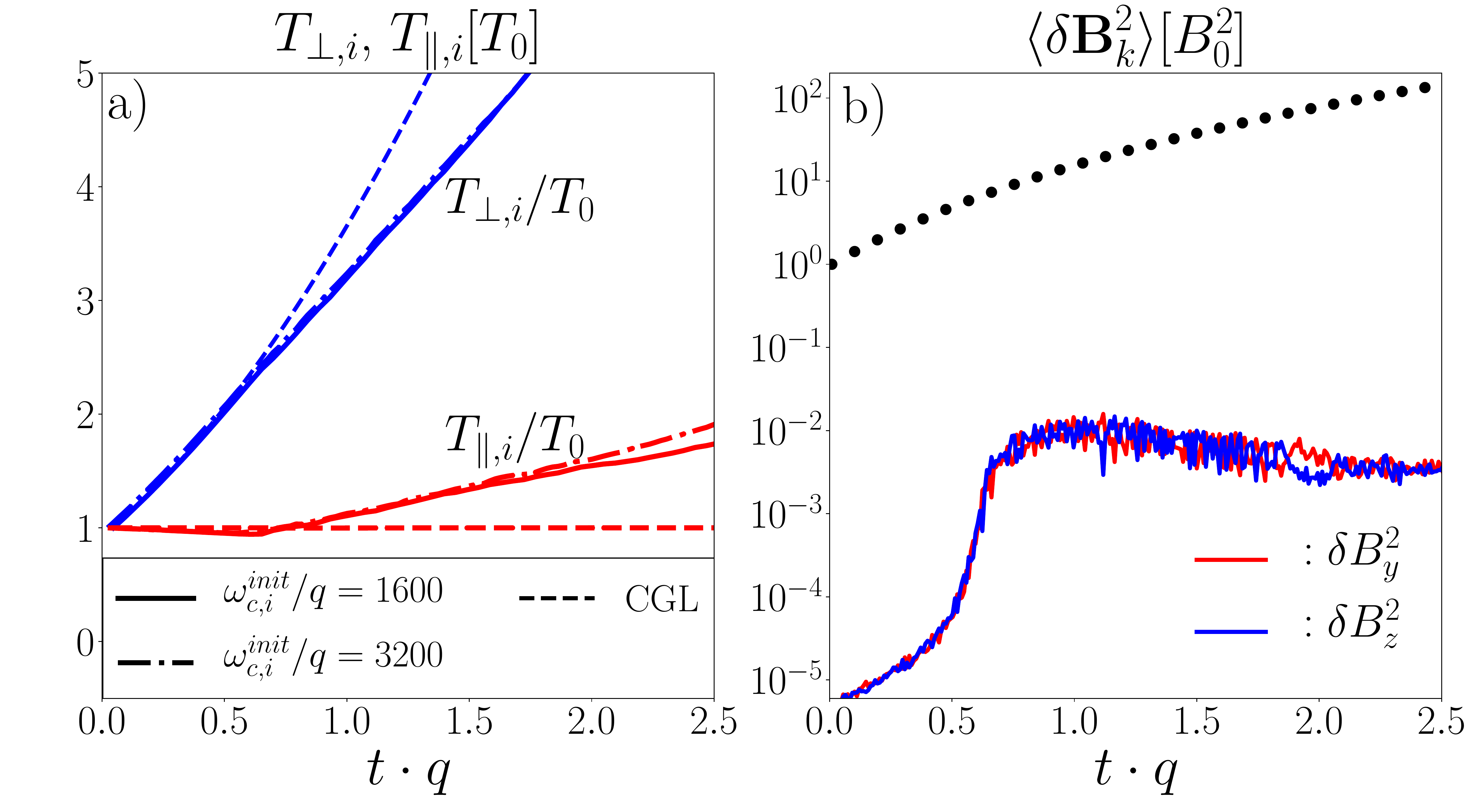}
}
\vspace*{-0.3cm}
\\
\subfloat{
  \centering
\hspace*{-0.5cm}
\includegraphics[width=0.495\textwidth]{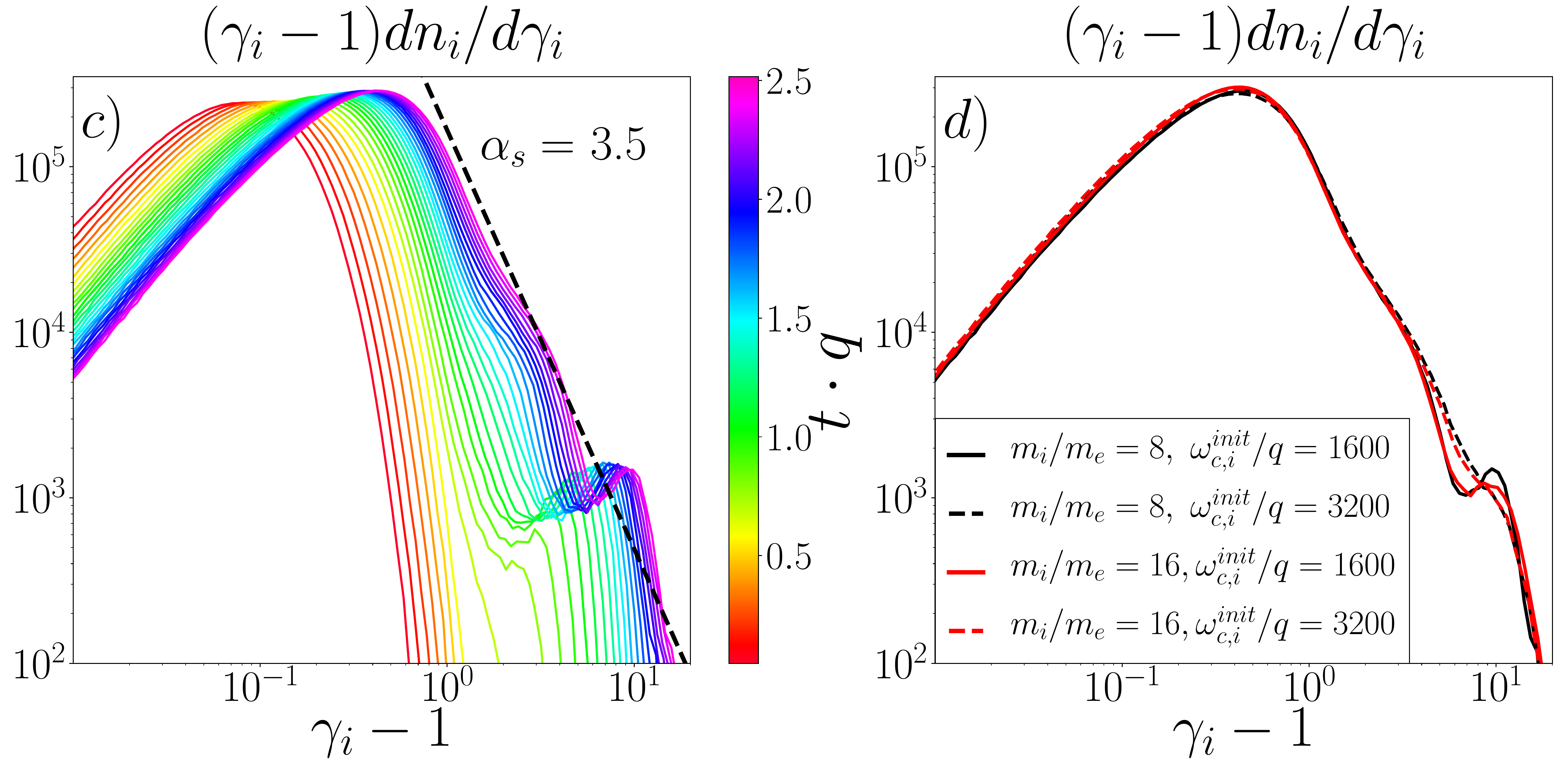}  
}
\vspace*{-0.2cm}
\caption{\small Panel $a$ shows the evolution of $T_{\perp,i}$ (blue) and $T_{\parallel,i}$ (red) for compressing runs C1m8b0.5a ($m_i/m_e=8$ and $\omega_{c,i}^{\textrm{init}}/q=1600$; solid) and C1m8b0.5b ($m_i/m_e=8$ and $\omega_{c,i}^{\textrm{init}}/q=3200$; dashed-dotted), both with $\beta_i^{\textrm{init}}=0.5$ and $k_BT_i=k_BT_e=0.05m_ic^2$. The double adiabatic predictions are shown in dashed lines \citep{ChewEtAl1956}. Panel $b$ shows $\delta B_y^2$ (red) and $\delta B_z^2$ (blue) normalized to $B_0^2$. The dotted line shows $|\langle \textbf{\textit{B}} \rangle|^2$. Panel $c$ shows the evolution of the ion spectrum of run C1m8b0.5a. Panel $d$ shows a comparison of the spectra at $t\cdot q=2.5$ for runs with $m_i/m_e=8$ and $\omega_{c,i}^{\textrm{init}}/q=1600$ (C1m8b0.5a; solid black),  $m_i/m_e=8$ and $\omega_{c,i}^{\textrm{init}}/q=3200$ (C1m8b0.5b; dashed black), $m_i/m_e=16$ and $\omega_{c,i}^{\textrm{init}}/q=1600$ (C1m16b0.5a; solid red) and $m_i/m_e=16$ and $\omega_{c,i}^{\textrm{init}}/q=3200$ (C1m16b0.5b; dashed red).}
\label{fig:compress}
\end{figure}

\noindent The compression runs use the same initial plasma parameters as the shearing runs that show significant ion acceleration: $\beta_i^{\textrm{init}}=0.5$, $k_BT_i/m_ic^2=0.05$, and $T_i=T_e$, which allows a direct comparison between the two setups. Fig. \ref{fig:compress}$a$ shows the evolution of $T_{\perp,i}$ (solid-blue) and $T_{\parallel,i}$ (solid-red) for run C1m8b0.5a, which uses $m_i/m_e=8$ and $\omega_{c,i}^{\textrm{init}}/q=1600$ ($k_BT_{\perp,i}=p_{\perp,i}/n_i$ and $k_BT_{\parallel,i}=p_{\parallel,i}/n_i$, where $n_i$ is the ion density).\footnote{We show $T_{\perp,i}$ and $T_{\parallel,i}$ instead of $p_{\perp,i}$ and $p_{\parallel,i}$ in order to disentangle variations in $n_i$ through the compression from the variation through double-adiabatic and instability-induced effects.} Both temperatures initially follow the double adiabatic evolution reasonably well, which is marked with the dashed-blue and dashed-red lines, respectively \citep{ChewEtAl1956}.\footnote{The small initial discrepancies ($t\cdot s \lesssim 0.6$) between the double-adiabatic predictions and $T_{\parallel,i}$ and $T_{\perp,i}$ are because these predictions assume that the ions are non-relativistic, which is mildly broken for our $k_BT_i/m_ic^2 \sim 0.1$ ions.} The adiabatic evolution breaks at $t\cdot q \approx 0.6$, after the exponential growth of the IC modes begins. This exponential growth and subsequent saturation can be seen from Fig. \ref{fig:compress}$b$, which shows the $\delta B_y^2$ and $\delta B_z^2$ components of run C1m8b0.5a, which evolve quite similarly to the $\delta B_{\perp,xy}$ and $\delta B_z^2$ fluctuations of analogous shearing runs (see, e.g., the cases of runs S2m2b0.5 and S2m10b0.5 in Fig. \ref{fig:evolb}$a$).\newline

\noindent The evolution of the ion spectrum of run C1m8b0.5a is shown in Fig. \ref{fig:compress}$c$. We see a remarkable similarity with the spectral evolution of the shearing runs with $\beta_i^{\textrm{init}}=0.5$, shown in Figs. \ref{fig:spectra}$a$ and \ref{1d2dspect}$a$. Indeed, also in the compressing case, there is a growth of a non-thermal tail that, by $t\cdot q=2.5$, can be described as a power law of index $\alpha \approx 3.5$ plus two bumps. \newline

\noindent We also test the dependence of this acceleration on both $m_i/m_e$ and ion magnetization $\omega_{c,i}^{\textrm{init}}/q$. Fig. \ref{fig:compress}$d$ shows a comparison between cases with $m_i/m_e=8$ and 16 (black and red, respectively), and with $\omega_{c,i}/q=1600$ and 3200 (solid and dashed, respectively), which share the same initial parameters: $\beta_i^{\textrm{init}}=0.5$, $k_BT_i/m_ic^2=0.05$ and $T_i=T_e$. We see essentially no difference between the runs with different values of $m_i/m_e$, and only a {\it slight} hardening of the tail as $\omega_{c,i}/q$ increases, in agreement with the shearing results of \S \ref{mimewcisdepend}. \newline

\noindent These results essentially reproduce our shearing runs, both in terms of the significance of the ion acceleration in the regime: $\beta_i^{\textrm{init}}=0.5$, $k_BT_i/m_ic^2=0.05$ and $T_i=T_e$, and in terms of the almost independence on $m_i/m_e$ and $\omega_{c,i}/q$. \newline

\noindent As in the shearing case, the independence on $m_i/m_e$ can be understood as due to the resonant nature of the ion-IC interaction. The independence on $\omega_{c,i}^{\textrm{init}}/q$ implies, similarly to the shearing case, that the effective ion scattering rate $\nu_{eff,i}$ has to be $\propto q$. This can indeed be inferred by applying Eq. \ref{prateek1} to the compressing runs. In this case, $\hat{\textbf{\textit{b}}}\hat{\textbf{\textit{b}}}:\nabla \textbf{\textit{v}}=0$, thus (using the continuity relation $\partial n_i/\partial t = -n_i\nabla \cdot \textbf{\textit{v}}$) $n_i\partial T_{\parallel,i}/\partial t = (2/3) \nu_{eff,i}\Delta p_i$. However, Fig. \ref{fig:compress}$a$ shows that, after the saturation of the IC waves, $\partial T_{\parallel,i}/\partial t \sim (q/2) T_{\parallel,i}$, which suggests:
\begin{equation}  
\nu_{eff,i} \approx \frac{3}{4}q (p_{\parallel,i}/\Delta p_i).
\label{nuq}
\end{equation}
The ratio $p_{\parallel,i}/\Delta p_i$ evolves fairly independent of $\omega_{c,i}^{\textrm{init}}/q$, as can be seen from Fig. \ref{fig:compress}$a$, which shows $T_{\perp,i}$ and $T_{\parallel,i}$ for runs C1m8b0.5a ($\omega_{c,i}^{\textrm{init}}/q=1600$; solid line) and C1m8b0.5b ($\omega_{c,i}^{\textrm{init}}/q=3200$; dot-dashed line). Therefore, Eq. \ref{nuq} implies that $\nu_{eff,i}\propto q$.\newline

\noindent The similarity between Eqs. \ref{nus} and \ref{nuq}, is consistent with the ion spectra being similar in the two setups, when comparing spectra at equal values of $ts$ and $tq$. Indeed, considering that in the shearing and compressing cases $p_{\parallel,i}/\Delta p_i\sim 1/2$ during most of the saturated IC regime (see Figs. \ref{fig:cgl}$a$ and \ref{fig:compress}$a$), at equal values of $ts$ and $tq$, the ions must have experienced a similar number of effective scatterings. Thus, if the properties of the IC modes are comparable, their acceleration effects by the end of the shearing and compressing simulations should also be comparable.\footnote{Notice, however, that, although the initial ion conditions in the shearing and compressing runs are the same, the final conditions are somewhat different. For instance, in the shearing runs with $\beta_i^{\textrm{init}}=0.5$ the final value of the parallel ion beta is $\beta_{\parallel,i}^{\textrm{final}}\sim 0.025$ (considering the evolution of $p_{\parallel,i}$ seen in Fig. \ref{fig:cgl}$a$ and the expected evolution of $B$), while in the analogous compressing runs $\beta_{\parallel,i}^{\textrm{final}}\sim 0.1$ (considering the evolution of $T_{\parallel,i}$ seen in Fig. \ref{fig:compress}$a$ and the expected evolutions of $n_i$ and $B$). This, plus the different factors on the right hand sides of Eqs. \ref{nus} and \ref{nuq}, imply that the final ion spectra in these two setups should be similar but not necessarily the same.}\newline

\section{Summary and Conclusions}
\label{sec:conclu}
\noindent Our 1D and 2D particle-in-cell (PIC) plasma simulations show that ions can be stochastically accelerated by the inelastic scattering provided by the ion-cyclotron (IC) instability. This acceleration occurs in the non-linear, saturated state of the instability, which is reached due to a permanent amplification of the background magnetic field $\langle \textbf{\textit{B}} \rangle$.\newline

\noindent In the regime in which initially $k_BT_i=k_BT_e=0.05m_ic^2$, we show that the IC ion acceleration is significant if $\beta_i^{\textrm{init}} \lesssim 1$. This is demonstrated by comparing 2D simulations with $\beta_i^{\textrm{init}}=0.5$ and 2. When $\beta_i^{\textrm{init}}=0.5$, the ion scattering is dominated by the IC instability, which produces a non-thermal tail in the ion energy spectrum. After \textbf{\textit{B}} is amplified by a factor $\sim 2.7$, the tail can be approximately described as a power-law of index $\sim 3.4$ plus two non-thermal bumps. The maximum ion Lorentz factor at that time is $\gamma_i \sim 10$, but it continues to grow at the end of the simulation. Also, the tail accounts for $2-3\%$ of the ions and $\sim 18\%$ of their kinetic energy. On the other hand, when $\beta_i^{\textrm{init}} =2$, the ion scattering is dominated by the mirror instability (the IC modes are subdominant) and the acceleration is significantly suppressed.\newline

\noindent In the IC dominated regime, as the ion scattering increases the energy of the ions of the tail, it reduces the energy of the ions in the thermal part of the spectrum (see Fig. \ref{fig:testp}). This is consistent with the IC modes being driven unstable mainly by the pressure anisotropy of the thermal ions. This way, the role of the IC modes is to absorb part of the energy of the thermal ions and give it to the non-thermal ions. This process is very similar to the stochastic acceleration of electrons by the whistler instability found by \cite{RiquelmeEtAl2017}. The efficiency of the ion acceleration, therefore, relies on the IC modes being able to provide resonant scattering to both thermal and non-thermal ions. We analyzed the consistency of this scenario by calculating the $k$ numbers and phase velocities of the dominant IC modes, showing that they can resonate with ions of all the energies (see discussion in \S \ref{sec:resonance}). \newline

\noindent Given that our simulations can not use realistic values of $m_i/m_e$ and $\omega_{c,i}^{\textrm{init}}/s$, one important aspect of our study is to ensure that these parameters do not affect the acceleration. Thus, first we ensure that the dominance of the IC modes for $\beta_i^{\textrm{init}}=0.5$ does not depend on $m_i/m_e$ and $\omega_{c,i}^{\textrm{init}}/s$. This was done comparing 2D simulations with $m_i/m_e=2$ and 10 (see \S \ref{sec:icregime}), and also using theoretical, linear dispersion relation calculations to determine the pressure anisotropy needed for the growth of the IC and mirror instabilities in astrophysically realistic conditions (see Appendix \S \ref{linear}). Both analyses show that, in realistic astrophysical plasmas, the IC instability dominates in the regime $\beta_i^{\textrm{init}} \lesssim 1$, at least for the case of $T_e=T_i$ explored here. \newline

\noindent Then, using that the dominant IC wave vectors $\textbf{\textit{k}}$ satisfy $\textbf{\textit{k}} \parallel \langle\textbf{\textit{B}}\rangle$, we use computationally cheaper 1D shear simulations to test the ion acceleration using a significantly larger range of values for $m_i/m_e$ and $\omega_{c,i}^{\textrm{init}}/s$. While $m_i/m_e$ almost does not affect the acceleration, increasing $\omega_{c,i}^{\textrm{init}}/s$ only produces a slight hardening of the non-thermal tail (see \S \ref{mimewcisdepend}). This almost complete independence of the acceleration on $\omega_{c,i}^{\textrm{init}}/s$ is consistent with the effective ion scattering rate $\nu_{eff,i}$ being proportional to $s$. This condition is indeed needed in order to have the continuous driving of the ion pressure anisotropy being nearly compensated by the pitch-angle scattering.\newline

\noindent In order to assess the importance of the specific large scale mechanism that amplifies the background magnetic field, we also ran compressing box PIC simulations like in \cite{SironiEtAl2015}. We find essentially no difference in the ion acceleration efficiency between the shearing and compressing cases (see \S \ref{sec:comp}).\newline

\noindent Our work is valid in a sub-relativistic regime in which initially $k_BT_i=k_BT_e=0.05m_ic^2$. This regime can be relevant in the inner region of low-luminosity accretion disks around black holes (where the collisionless plasma condition is expected). In these systems, the condition $\beta_i^{\textrm{init}} \lesssim 1$ required for the acceleration is most likely satisfied in the coronal region of the disks \citep[e.g.,][]{ChaelEtAl2018}. \newline

\noindent Nevertheless, assessing the importance of the presented acceleration mechanism in these and other astrophysical systems requires a more complete understanding of its dependence on plasma parameters, as well as clarifying the importance of possible long term evolution effects. Indeed, in this work we focus on a single value of $k_BT_i$ and use $T_e/T_i=1$. However, varying these parameters may affect the IC physics and, therefore, the efficiency of the ion acceleration. For instance, having $T_e/T_i \ll 1$ may increase significantly the values of $\beta_i^{\textrm{init}}$ for which the IC instability dominates \citep{SironiEtAl2015, Sironi2015}. Since the condition $T_e/T_i \ll 1$ is most likely satisfied in low luminosity disks \citep{NarayanEtAl1995,YuanEtAl2003}, this could increase the importance of the presented ion acceleration in these systems. \newline

\noindent In terms of the long term evolution of the acceleration process, in the turbulent environment of accretion disks we expect many successive $\delta B/B\sim 1$ amplifications and decreases of the field \cite[see also][]{VerscharenEtAl2016}. So a more realistic picture of this process should consider the acceleration presented in this work occurring many times as the plasma is gradually accreted. We will study these aspects of the acceleration process in future investigations.  \newline 
 
\acknowledgements
\noindent This research was supported by the supercomputing infrastructure of the NLHPC (ECM-02) at the Center for Mathematical Modeling of University of Chile, by the Habanero cluster at Columbia University, and by the XSEDE computing system (allocations TG-AST140039, TG-AST140083 and TG-PHY160040). F.L. acknowledges support from the NSF grant NST AST-1616037. L.S. thanks support from DoE DE-SC0016542, NASA ATP NNX-17AG21G, NSF ACI1657507, and NSF AST-1716567. D.V. was supported by the STFC Ernest Rutherford Fellowship ST/P/003826/1.

\appendix
\section{{\bf A.} 1D shear setup}
\label{1druns}
\noindent The goal of our 1D shear runs is to simulate a shearing plasma assuming that its properties depend only on the direction parallel to the mean background magnetic field $\langle \textbf{\textit{B}} \rangle$. This is equivalent to assuming that the wave vectors $\textbf{\textit{k}}$ captured in the simulations satisfy $\textbf{\textit{k}}\parallel \langle \textbf{\textit{B}} \rangle$. Since in a shearing plasma the direction of $\langle \textbf{\textit{B}} \rangle$ evolves with time, the orientation of the wave vectors that can be consistently resolved needs to evolve accordingly.\newline 

\noindent Our 1D setup is built upon the ``shearing coordinates" setup presented by \cite{RiquelmeEtAl2012}. This setup was designed so that the simulation domain follows the shearing flow of the plasma, which is given by the shear velocity $\textbf{\textit{v}}$ ($=v\hat{y}$). Figures \ref{fig:shearscheme}$a$ and \ref{fig:shearscheme}$b$ illustrate the way the shape of the domain evolves as seen by an inertial observer at $t=0$ and $t>0$, respectively. Formally, the shearing coordinates $(x',y',z',t')$ are defined in terms of the regular, inertial coordinates $(x, y, z, t)$ as 
\begin{equation}
x'=x, \hspace{3em} y'=\Gamma(y-vt), \hspace{3em} z'=z, \hspace{3em} \textrm{and} \hspace{3em} t'=\Gamma(t-vy/c^2),
\label{r12}
\end{equation}
where $\Gamma=(1-v^2/c^2)^{-1/2}$, $v=-c\textrm{ arctanh}(sx/c)$,\footnote{Notice that, in the limit $|v| \ll c$, the $v=-c\textrm{ arctanh}(sx/c)$ expression is equivalent to $v=-sx$, as we assume in Figures \ref{fig:shearscheme}$a$ and \ref{fig:shearscheme}$b$. Since $v=-c\textrm{ arctanh}(sx/c)$ ensures that $|v| < c$, this expression was adopted in \cite{RiquelmeEtAl2012} in order to deduce the equations that describe the dynamics of the plasma in the shearing coordinate system. However, both in \cite{RiquelmeEtAl2012} and in this paper we are interested in the plasma dynamics in the regime $|v| \ll c$.} $c$ is the speed of light and $s$ is the shear rate of the plasma. \newline

\noindent Our 1D domain is defined by $y'=0$, and it is shown by the blue lines in Figures \ref{fig:shearscheme}$a$ and \ref{fig:shearscheme}$b$. These figures show how the length and orientation of the 1D domain change over time. This in turn changes the orientation of the wave vectors $\textbf{\textit{k}}$ that can be captured within the domain. In addition, there is an initially homogeneous magnetic field, $\textbf{\textit{B}}$, pointing parallel to the 1D domain at $t=0$. Magnetic flux conservation ensures that $\langle \textbf{\textit{B}} \rangle$ will always be parallel to the 1D domain at $t>0$, as also depicted in Figures \ref{fig:shearscheme}$a$ and \ref{fig:shearscheme}$b$. Therefore, if the problem of interest is dominated by waves that propagate parallel to $\langle \textbf{\textit{B}} \rangle$, our 1D simulations will be able to capture the essence of the phenomenon.\newline

\noindent In order for our 1D simulations to consistently satisfy $\textbf{\textit{k}} \parallel \langle \textbf{\textit{B}} \rangle$, we need to replace $x'$ with a new coordinate $x'_1$, which we define as 
\begin{equation}
\label{trans}
x'_1\equiv x'-\frac{y'st'}{1+s^2t'^2}.
\end{equation}
Indeed, if we assume $|v| \ll c$ and $sy' \ll c$, the partial derivatives with respect to the spatial coordinates in the inertial frame are
\begin{equation}
\frac{\partial}{\partial x} = \frac{1}{1+s^2t^2}\frac{\partial}{\partial x'_1} + st\frac{\partial}{\partial y'}, \hspace{3em} \frac{\partial}{\partial y} = \frac{-st}{1+s^2t^2}\frac{\partial}{\partial x'_1} + \frac{\partial}{\partial y'}, \hspace{3em} \textrm{and} \hspace{3em} \frac{\partial}{\partial z} = \frac{\partial}{\partial z'}.
\label{part}
\end{equation}
Thus, if we impose the 1D condition, namely that the fields depend only on $x'_1$, with $\partial/\partial y' = \partial/\partial z' = 0$, Equations \ref{part} imply that the gradient of any field component in the inertial frame will have coordinates proportional to the vector ($1, -st, 0$). In the inertial frame the background magnetic field $\langle \textbf{\textit{B}} \rangle$ has components ($B_0, 0, 0$) at $t=0$, then, due to magnetic flux freezing, at $t>0$ these components will be $B_0$($1, -st, 0$). Thus, after imposing $\partial/\partial y' = \partial/\partial z' = 0$, our 1D simulations are able to capture all the modes with wave vectors $\textbf{\textit{k}}$ parallel to $\langle \textbf{\textit{B}}\rangle$, despite the fact that $\langle \textbf{\textit{B}}\rangle$ changes orientation over time. \newline

\noindent The assumptions $|v| \ll c$ and $sy' \ll c$ are equivalent to assuming that the plasma region of interest has a typical size, $L$, that satisfies $L \ll c/s$. For the study of kinetic instabilities, $L$ is typically of the order of the Larmor radius of the particles. Thus, the restriction $L \ll c/s$ applied to species $j$ becomes: $[\gamma_js/\omega_{c,j}][v_{j,\perp}/c]\ll 1$, where $\gamma_j$, $\omega_{c,j}$, and $v_{j,\perp}$ are the Lorentz factor, (non-relativistic) cyclotron frequency, and velocity perpendicular to $\textbf{\textit{B}}$ of particles $j$, respectively. In typical astrophysical environments, $\omega_{c,j}$ is many orders of magnitude larger than $s$, therefore the assumption $L \ll c/s$ is reasonable for the study of kinetic instabilities.\newline

\noindent In what follows we will give expressions for the evolution of fields and particle momenta and positions in our 1D shear setup. The condition $L \ll c/s$ is assumed in all of these expressions.     

\subsection{Evolution of $\textbf{\textit{E}}'$ and $\textbf{\textit{B}}'$}

\noindent Considering the evolutions of $\textbf{\textit{E}}'$ and $\textbf{\textit{B}}'$ in terms of $\textbf{\textit{J}}'$ in the shearing coordinates presented in Equations A14 and A26 of \cite{RiquelmeEtAl2012}, after replacing $x'$ with $x'_1$ (Eq. \ref{trans}) and assuming $\partial/\partial y' = \partial/\partial z' = 0$, the fields dynamics is given by:
\begin{equation}
\frac{\partial B_x'}{\partial t'} = c\frac{st'}{1+s^2t'^2} \frac{\partial E_z'}{\partial x'_1}, \hspace{1em} \frac{\partial B_y'}{\partial t'} = c\frac{1}{1+s^2t'^2} \frac{\partial E_z'}{\partial x'_1} - sB_x', \hspace{1em} \frac{\partial B_z'}{\partial t'} = \frac{-c}{1+s^2t'^2}\Big( \frac{\partial E_y'}{\partial x'_1} + st'\frac{\partial E_x'}{\partial x'_1}\Big),
\label{maxb}
\end{equation}
and
\begin{equation}
\frac{\partial E_x'}{\partial t'} = -c\frac{st'}{1+s^2t'^2} \frac{\partial B_z'}{\partial x'_1}-4\pi J'_x, \hspace{1em} \frac{\partial E_y'}{\partial t'} = -c\frac{1}{1+s^2t'^2} \frac{\partial B_z'}{\partial x'_1} - sE_x'-4\pi J'_y, \hspace{1em} \frac{\partial E_z'}{\partial t'} = \frac{c}{1+s^2t'^2}\Big(\frac{\partial B_y'}{\partial x'_1} + st'\frac{\partial B_x'}{\partial x'_1}\Big) -4\pi J'_z.
\label{maxe}
\end{equation}
In Equations \ref{maxb} and \ref{maxe}, $\textbf{\textit{E}}'$, $\textbf{\textit{B}}'$, and $\textbf{\textit{J}}'$, are defined according to the conventional relativistic transformation of the electric field, magnetic field, and current density from the inertial frame to the frame of an observer that moves with velocity $\textbf{\textit{v}}$ (Eqs. A2 and A28 of \cite{RiquelmeEtAl2012}). This means that $\textbf{\textit{J}}'$ can be calculated directly from the motion of the particles in the shear coordinates, assuming that their momenta $\textbf{\textit{p}}'$ can be obtained from their momenta in the inertial frame $\textbf{\textit{p}}$, using the conventional relativistic momentum transformation (see Eq. A30 of \cite{RiquelmeEtAl2012}). 

\subsection{Evolution of particle momenta}
\noindent Since our 1D setup uses the same definitions of time ($t'$) and of particle momenta ($\textbf{\textit{p}}'$) as in the shearing coordinate system of \cite{RiquelmeEtAl2012}, the evolution of $\textbf{\textit{p}}'$ will not change and will be given by
\begin{equation}
\frac{d\textbf{\textit{p}}'}{dt'} = q_c\big(\textbf{\textit{E}}' + \frac{\textbf{\textit{u}}'}{c}\times \textbf{\textit{B}}'\big) + sp_x'\hat{y},
\label{mom}
\end{equation}
where $q_c$ is the particles' electric charge and $\textbf{\textit{u}}'=\textbf{\textit{p}}'/m\gamma'$, with $m$ and $\gamma'$ being the particle mass and Lorentz factor, respectively. Eq. \ref{mom} corresponds to Eqs. A31 of \cite{RiquelmeEtAl2012} in the limit $L \ll c/s$. 

\subsection{Evolution of particle positions}  
\noindent By taking the derivative of Equation \ref{trans} with respect to $t'$, and applying the limit $L \ll c/s$ to the expressions for $dx'/dt'$ and $dy'/dt'$ given by Eq. A35 of \cite{RiquelmeEtAl2012}, we find  
\begin{equation}
\frac{dx'_1}{dt'} = u_x'\frac{1}{1+s^2t'^2} - u_y'\frac{st'}{1+s^2t'^2},
\label{posevol1}
\end{equation} 
where $u_x'$ and $u_y'$ are the $x$ and $y$ components of $\textbf{\textit{u}}'$. Eq. \ref{posevol1} implies that $dx'_1/dt'$ is the scalar product between $\textbf{\textit{u}}'$ and $\langle \textbf{\textit{B}} \rangle/|\langle \textbf{\textit{B}} \rangle|$ ($=(\hat{x} - st'\hat{y})/(1+s^2t'^2)^{1/2}$), corrected by the ``expansion" of the 1D domain (represented by the $y'=0$ region depicted in blue in Figures \ref{fig:shearscheme}$a$ and \ref{fig:shearscheme}$b$), which provides the extra common factor $1/(1+s^2t'^2)^{1/2}$ on the right hand side of Eq. \ref{posevol1}. This implies that $x'_1$ correctly represents particle displacements along $\langle \textbf{\textit{B}} \rangle$.

\subsection{Charge conservation}

\noindent The evolution of particle positions in our simulations is consistent with the conservation of electric charge density. This can be checked by considering Gauss' law in the inertial frame: $\nabla \cdot \textbf{\textit{E}}=4\pi \rho_c$, where $\textbf{\textit{E}}$ and $\rho_c$ are the electric field and charge density in that frame. Indeed, applying the transformations given by Eqs. \ref{r12} to Gauss' law in the limit $L \ll c/s$, and assuming first the more general problem in 3D described in \cite{RiquelmeEtAl2012}:
\begin{equation}
\frac{\partial \rho_c'}{\partial t'} + \frac{\partial J_x'}{\partial x'} + \frac{\partial (J_y'+st'J_x') }{\partial y'} + \frac{\partial J_z'}{\partial z'}=0,
\label{conserve2}
\end{equation}
where $\rho_c'$ is the charge density of the plasma calculated using the shearing coordinates, which, as shown in Eq. A29 of \cite{RiquelmeEtAl2012}, is equal to $\rho_c$ in the limit $L \ll c/s$. Thus, according to Eq. \ref{conserve2}, in order to satisfy charge conservation in our multidimensional shearing coordinate, particle motions in $y'$ lead to an effective current along the $y'$ direction of $J'_y+st'J'_x$. Thus, charge conservation is ensured in our multidimensional shearing coordinate simulations by evolving $x', y'$ and $z'$ according to

\begin{equation}
\frac{dx'}{dt'}=u_x', \hspace{1em} \frac{dy'}{dt'}=u_y'+st'u_x', \hspace{1em} \textrm{and} \hspace{1em}\frac{dz'}{dt'}=u_z',
\label{dy}
\end{equation}
which we do in the 2D simulations presented in this paper and in our previous works \citep[e.g.,][]{RiquelmeEtAl2017}. Eq. \ref{dy} is equivalent to Eq. A35 of \cite{RiquelmeEtAl2012} in the limit $L \ll c/s$.\footnote{Notice that the notation used in Eq. A35 of \cite{RiquelmeEtAl2012} is different from the one used in this work, and that we interprete their $u_x', u_y'$ and $u_z'$ as our $dx'/dt'$, $dy'/dt'$ and $dz'/dt'$.}\newline

\noindent Finally, by making an analogous analysis in the case of the 1D setup, we show that charge conservation implies
\begin{equation}
\frac{\partial \rho_c'}{\partial t'} + \frac{\partial}{\partial x'_1}\Big(\frac{J_x'}{1+s^2t'^2} - \frac{st'J_y'}{1+s^2t'^2}\Big)=0.
\label{conserve1}
\end{equation}
Thus, according to Eq. \ref{conserve1}, in order for our 1D runs to conserve charge, $x'_1$ must evolve so that the effective current along $\langle \textbf{\textit{B}} \rangle$ is $(J_x'-st'J_y')/(1+s^2t'^2)$, which is indeed satisfied by the evolution of $x_1'$ given by Eq. \ref{posevol1}. \newline

\section{{\bf B.} IC vs. mirror under realistic conditions: linear theory analysis}
\label{linear}
\begin{figure}[t!]
\hspace*{-0.7cm}
\vspace*{-0.5cm}
\center
\includegraphics[width=.46\textwidth]{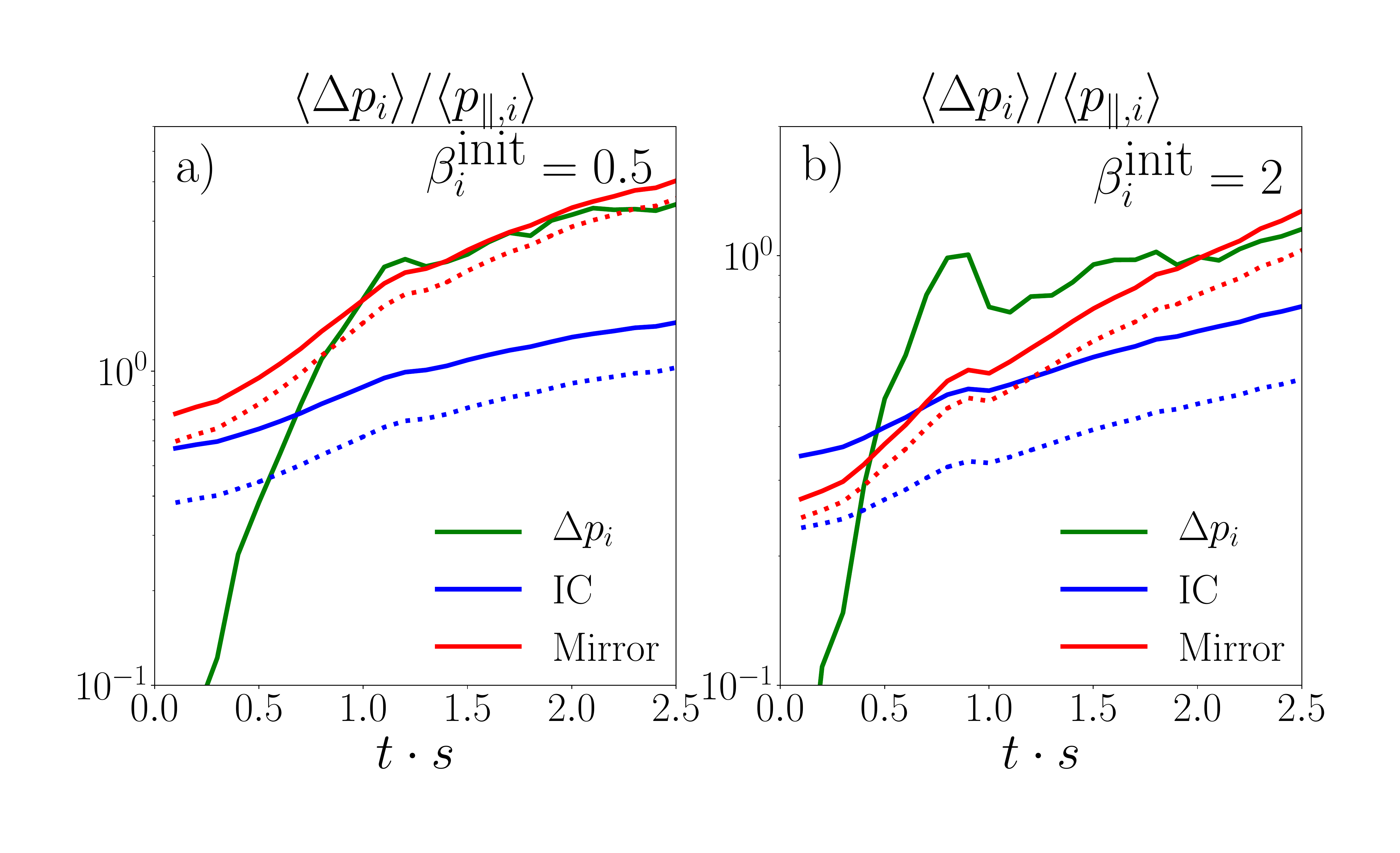}
\vspace*{-0.5cm}
\caption{\small Panels $a$ and $b$: the mirror (solid red) and IC (solid blue) linear anisotropy thresholds (assuming growth rate $s$) in 2D runs S2m2b0.5 ($m_i/m_e=2$ and $\beta_i^{\textrm{init}}=0.5$; panel $a$) and S2m2b2 ($m_i/m_e=2$ and $\beta_i^{\textrm{init}}=2$; panel $b$). The anisotropies obtained directly from the simulations are in solid green. The dotted lines show analogous linear anisotropy thresholds for the mirror (dotted red) and IC (dotted blue) instabilities assuming $m_i/m_e=1836$ and a growth rate equal to $10^{-6}\omega_{c,i}$.}
\vspace*{0.2cm}
\label{fig:evolp}
\end{figure}
\noindent In order to inquire whether the dominance of the IC instability for $\beta_i^{\textrm{init}} \lesssim 1$ continues to be valid in realistic, astrophysically relevant regimes ($m_i/m_e=1836$ and $\omega_{c,i}/s \gg 800$, which we can not study with our 2D simulations), we make use of linear theory. Thus, we calculate the threshold ion pressure anisotropy, $\Delta p_i/p_{\parallel,i} \equiv (p_{\perp,i}-p_{\parallel,i})/p_{\parallel,i}$, needed for the growth of IC and mirror modes at a rate $s$, using the linear Vlasov solver NHDS \citep[][which assumes bi-Maxwellian, non-relativistic ion velocity distributions]{VerscharenEtAl2018}.\footnote{Modern linear solvers to the linear Vlasov-Maxwell system of equations can account for relativistic effects and non-Maxwellian background distributions \citep{VerscharenEtAl2018b}. Studies using this type of solver can evaluate these effects; however, their application is beyond the scope of this paper.} This condition is motivated by the assumption that, in order to maintain $\Delta p_i/p_{\parallel,i}$ at a nearly stationary level (as we see in Figs. \ref{fig:evolp}$a$ and \ref{fig:evolp}$b$), the modes that provide the pitch-angle scattering must grow at roughly the rate at which the anisotropy is driven, which is $\sim s$. Thus, if non-linear effects did not play any significant role, the dominant instability should be the one with the lowest theoretical anisotropy threshold for a given value of $s$.\newline

\noindent However, non-linear effects are expected to be important and to affect the IC and mirror instabilities differently. Indeed, in-situ observations in the solar wind show significant discrepancies between linear theory and the measured ion anisotropy in regions of parameter space in which the IC instability should dominate \citep{HellingerEtAl2006,BaleEtAl2009}. One possible explanation is the departure from bi-Maxwellian ion velocity distributions observed in the solar wind, which may affect the efficiency of the resonant scattering between ions and the IC modes \citep{IsenbergEtAl2013}. Another possibility is the inhomogeneity in the magnetic field produced by the (subdominant but still present) mirror modes, which may also affect this resonance \citep{SouthwoodEtAl1993}. Our approach is thus to estimate these non-linear effects using the simulated cases. These simulations thus provide us with a calibration of the linear theory criterion for determining the dominant instability, which can then be applied to astrophysically realistic regimes.\newline

\noindent Figures \ref{fig:evolp}$a$ and \ref{fig:evolp}$b$ show the linear anisotropy threshold given by the mirror (solid red) and IC (solid blue) instabilities with growth rate $s$ in runs S2m2b0.5 and S2m2b2, which have $\beta_i^{\textrm{init}}=0.5$ and 2, respectively (both with $m_i/m_e=2$ and $\omega_{c,i}^{\textrm{init}}/s=800$), and compare them with the anisotropies obtained from the simulations (solid green). First, in the IC dominated regime (run S2m2b0.5), the IC threshold is $\sim 3$ times smaller than the ion anisotropy obtained from the simulation, showing that, in the case of the IC instability, non-linear effects give rise to ion anisotropies significantly larger than what is implied by the linear theory threshold. Also, the IC threshold is at least $\sim 2$ times smaller than the mirror threshold (Fig. \ref{fig:evolp}$a$), while in the mirror dominated case (Fig. \ref{fig:evolp}$b$) the IC threshold is at most $\sim 1.5$ times smaller. These results suggest that, in order for the IC instability to dominate, the IC threshold should be at least $\sim 2$ times smaller than the mirror threshold.\newline

\noindent We thus apply this criterion to astrophysically realistic cases, in which $m_i/m_e=1836$ and the instabilities grow at a rate of $10^{-6}\omega_{c,i}$. Figures \ref{fig:evolp}$a$ and \ref{fig:evolp}$b$ show in dotted lines the corresponding linear anisotropy thresholds for the mirror (dotted red) and IC (dotted blue) instabilities for $\beta_i^{\textrm{init}}=0.5$ and 2, respectively. In the $\beta_i^{\textrm{init}}=0.5$ case, the IC threshold is always at least $\sim 3$ times smaller than the mirror threshold, implying that, under realistic conditions, the IC modes continue to dominate in the $\beta_i^{\textrm{init}}=0.5$ case. In the $\beta_i^{\textrm{init}}=2$, on the other hand, the linear IC threshold is smaller than the mirror threshold by a factor $\lesssim 2$ during most of the simulations. Since we estimate that, in order for the IC instability to dominate, the mirror anisotropy threshold should be at least $\sim 2$ times larger than the one of the IC modes, this suggests that the mirror instability continues to dominate in this case.\newline

\begin{thebibliography}{99}
\addcontentsline{toc}{section}{Bibliography}

\bibitem[Anderson et al. (1991)]{AndersonEtAl1991} Anderson, B. J., Fuselier, S. A., \& Murr, D. 1991, GeoRL, 18, 1955
\bibitem[Bale et al. (2009)]{BaleEtAl2009} Bale, S. D.,  Kasper, J. C., Howes, G. G., Quataert, E., Salem, C., \& Sundkvist, D. 2009, PRL 103, 211101
\bibitem[Buneman(1993)]{Buneman93} Buneman, O. 1993, ``Computer Space Plasma Physics'', Terra Scientific, Tokyo, 67
\bibitem[Chael et al. (2018)]{ChaelEtAl2018} Chael, A, Rowan, M, Narayan, R., Johnson, M. \& Sironi, L. 2018, MNRAS, 478, 5209
\bibitem[Chandran (2003)]{Chandran2003} Chandran, B. 2003, \apj,  599, 1426.
\bibitem[Chew et al. (1956)]{ChewEtAl1956} Chew, G. F., Goldberger, M. L., \& Low, F. E. 1956, RSPSA, 236, 112
\bibitem[Cho \& Lazarian (2006)]{ChoEtAl2006} Cho, J. \& Lazarian, A. 2006, \apj, 638, 811
\bibitem[Dermer et al. (1996)]{DermerEtAl1996} Dermer, C. D., Miller, J. A. \& Li, H. 1996, \apj, 456, 106
\bibitem[Gary (1992)]{Gary1992} Gary, S. P. 1992, JGR, 97, 8519
\bibitem[Gary \& Wang (1996)]{GaryEtAl1996} Gary, S. P. \& Wang, J. 1996, J. Geophys. Res., 101, 10749
\bibitem[Hasegawa (1969)]{Hasegawa1969} Hasegawa, A. 1969, PhFl, 12, 2642
\bibitem[Hellinger et al. (2006)]{HellingerEtAl2006} Hellinger, P., Travnicek, P.,  Kasper, J. C., \&  Lazarus, A. J. 2006, GRL, 33, L09101
\bibitem[Isenberg et al. (2013)]{IsenbergEtAl2013} Isenberg, P. A., Maruca, B. A., \& Kasper, J. C. 2013, ApJ, 773, 164
\bibitem[Kulsrud \& Pearce (1969)]{KulsrudEtAl1969} Kulsrud, R. M. \& Pearce, W. P. 1969, \apj, 156, 445
\bibitem[Kulsrud(1983)]{Kulsrud1983} Kulsrud, R. M. 1983, in Handbook of Plasma Physics, ed. M. N. Rosenbluth \& R. Z. Sagdeev (Amsterdam: North Holland), 115
\bibitem[Lynn et al.(2014)]{LynnEtAl2014} Lynn, J., Quataert, E., \& Chandran, B., \& Parrish, I. 2014, 791, 71
\bibitem[Lyutikov (2007)]{Lyutikov2007} Lyutikov, M. 2007, ApJL, 668, L1
\bibitem[Maruca et al.(2011)]{MarucaEtAl2011} Maruca, B. A., Kasper, J. C., \& Bale, S. D. 2011, PhRvL, 107, 201101
\bibitem[Narayan \& Yi (1995)]{NarayanEtAl1995} Narayan, R. \& Yi, I. 1995, \apj, 452, 710
\bibitem[Petrosian \& Liu (2004)]{PetrosianEtAl2004} Petrosian, V. \& Liu, S. 2004, \apj, 610, 550
\bibitem[Pokhotelov et al.(2004)]{PokhotelovEtAl2004} Pokhotelov, O. A., Sagdeev,R. Z., Balikhin, M. A.,\& Treuman,R.A. 2004, JGR, 109, A09213
\bibitem[Ponti et al.(2017)]{PontiEtAl2017} Ponti, G.,  George, E.,  Scaringi, S., Zhang, S., Jin, C., et al. 2017, \mnras, 468, 2447
\bibitem[Riquelme et al.(2012)]{RiquelmeEtAl2012} Riquelme, M.~A., Quataert, E., Sharma, P., \& Spitkovsky, A.\ 2012, \apj, 755, 50 
\bibitem[Riquelme et al.(2017)]{RiquelmeEtAl2017} Riquelme, M. A., Osorio, A., \& Quataert, E. 2017, \apj, 850, 113
\bibitem[Schekochihin et al.(2005)]{SchekochihinEtAl2005} Schekochihin, A. A., Cowley, S. C., Kulsrud, R. M., Hammett, G. W., \& Sharma, P. 2005, ApJ, 629, 139
\bibitem[Sharma et al.(2006)]{SharmaEtAl2006} Sharma, P., Hammett, G. W., Quataert, E., \& Stone, J. 2006, ApJ, 637, 952
\bibitem[Sharma et al.(2007)]{SharmaEtAl2007} Sharma, P., Quataert, E., Hammett, G. W., \& Stone, J. 2007, \apj, 667, 714
\bibitem[Sironi \& Narayan (2015)]{SironiEtAl2015} Sironi, L., \& Narayan, R. 2015, ApJ, 800, 88
\bibitem[Sironi (2015)]{Sironi2015} Sironi, L. 2015, ApJ, 800, 89
\bibitem[Snyder et al.(1997)]{SnyderEtAl1997} Snyder, P. B., Hammett, G. W., \& Dorland, W. 1997, Phys. Plasmas, 4, 3974 
\bibitem[Southwood \& Kivelson (1993)]{SouthwoodEtAl1993} Southwood, D. J., \& Kivelson, M. G. 1993, JGR, 98, 9181
\bibitem[Spitkovsky(2005)]{Spitkovsky05} Spitkovsky, A. 2005, AIP Conf. Proc, 801, 345, astro-ph/0603211
\bibitem[Verscharen at al.(2016)]{VerscharenEtAl2016} Verscharen, D., Chandran, B., Klein, K. G., \& Quataert, E. 2016, 831, 128
\bibitem[Verscharen \& Chandran (2018)]{VerscharenEtAl2018} Verscharen, D. \& Chandran, B. 2018, Res. Notes AAS 2, 13
\bibitem[Verscharen et al. (2018)]{VerscharenEtAl2018b} Verscharen, D., Klein, K. G., Chandran, B., Stevens, M. L., Salem, C. S., \& Bale, S. D. 2018, J. Plasma Phys., 84, 905840403
\bibitem[Verscharen et al. (2019)]{VerscharenEtAl2019} Verscharen, D., Klein, K. G., \& Maruca, B. A. 2019, arXiv:1902.03448
\bibitem[Yuan et al.(2003)]{YuanEtAl2003} Yuan, F., Quataert, E., \& Narayan, R. 2003, \apj, 598, 301
\end{thebibliography}
\end{document}